\documentclass{aa}
\usepackage{graphicx}
\usepackage{xcolor}
\usepackage{float}
\usepackage{rotating}
\usepackage{booktabs}
\usepackage{mathtools}
\usepackage{placeins}
\usepackage{bm}
\usepackage{longtable}
\usepackage{adjustbox}
\usepackage{supertabular}
\usepackage{siunitx}
\usepackage{multirow}
\usepackage{tabularx}
\usepackage[switch]{lineno}

\usepackage{natbib}
\usepackage[breaklinks=True]{hyperref}
\usepackage{url}
\hypersetup{colorlinks=true,
            linkcolor=blue,
            citecolor=blue,
            filecolor=magenta,
            urlcolor=magenta}
            
\usepackage{footmisc}

\usepackage{graphicx}
\usepackage{txfonts}

\usepackage{twoopt}
\bibpunct{(}{)}{;}{a}{}{,}             
\makeatletter
  \newcommandtwoopt{\citeads}[3][][]{\href{http://adsabs.harvard.edu/abs/#3}%
    {\def\hyper@linkstart##1##2{}%
     \let\hyper@linkend\@empty\citealp[#1][#2]{#3}}}
  \newcommandtwoopt{\citepads}[3][][]{\href{http://adsabs.harvard.edu/abs/#3}%
    {\def\hyper@linkstart##1##2{}%
     \let\hyper@linkend\@empty\citep[#1][#2]{#3}}}
  \newcommandtwoopt{\citetads}[3][][]{\href{http://adsabs.harvard.edu/abs/#3}%
    {\def\hyper@linkstart##1##2{}%
     \let\hyper@linkend\@empty\citet[#1][#2]{#3}}}
  \newcommandtwoopt{\citeyearads}[3][][]%
    {\href{http://adsabs.harvard.edu/abs/#3}
    {\def\hyper@linkstart##1##2{}%
     \let\hyper@linkend\@empty\citeyear[#1][#2]{#3}}}
\makeatother
\begin{document}

   \title{Transiting exoplanets with the Mid-InfraRed Instrument on board the James Webb Space Telescope: From simulations to observations}

   \author{A. Dyrek
          \inst{1}
          \and
          E. Ducrot \inst{1}\thanks{Paris Region Fellow, Marie Sklodowska-Curie Action}
          \and
          P-O. Lagage \inst{1}
          \and
          P. Tremblin \inst{2}
          \and
          S. Kendrew \inst{3}
          \and
          J. Bouwman \inst{4}
          \and
          R. Bouffet \inst{1}
          }

   \institute{Université Paris Cité, Université Paris-Saclay, CEA, CNRS, AIM, F-91191, Gif-sur-Yvette, France 
    \and
    Université Paris-Saclay, UVSQ, CNRS, CEA, Maison de la Simulation, 91191, Gif-sur-Yvette, France
    \and
    European Space Agency, Space Telescope Science Institute, 3700 San Martin Dr., Baltimore, MD 21218, USA
    \and
    Max Planck Institute for Astronomy (MPIA), Königstuhl 17, D-69117 Heidelberg, Germany
             }


 
  \abstract
   {The James Webb Space Telescope (JWST) has now started its exploration of exoplanetary worlds. In particular, the Mid-InfraRed Instrument (MIRI) with its Low-Resolution Spectrometer (LRS) carries out transit, eclipse, and phase-curve spectroscopy of exoplanetary atmospheres with an unprecedented precision in a so far almost uncharted wavelength range.}
   {The precision and significance in the detection of molecules in exoplanetary atmospheres relies on a thorough understanding of the instrument itself and on accurate data reduction methods. This paper aims to provide a clear description of the instrumental systematics that affect observations of transiting exoplanets through the use of simulations.}
   {We carried out realistic simulations of transiting-exoplanet observations with the MIRI LRS instrument that included the model of the exoplanet system, the optical path of the telescope, the MIRI detector performances, and instrumental systematics and drifts that could alter the atmospheric features we are meant to detect in the data. After we introduce our pipeline, we show its performance on the transit of L168-9b, a super-Earth-sized exoplanet observed during the commissioning of the MIRI instrument.}
   {This paper provides a better understanding of the data themselves and of the best practices in terms of reduction and analysis through comparisons between simulations and real data. We show that simulations validate the current data-analysis methods. Simulations also highlight instrumental effects that impact the accuracy of our current spectral extraction techniques. These simulations are proven to be essential in the preparation of JWST observation programs and help us to assess the detectability of various atmospheric and surface scenarios.}
   {}

    \titlerunning{Transiting exoplanets with JWST MIRI: From simulations to observations}
    \maketitle
%

\section{Introduction}
    
The long-awaited James Webb Space Telescope (JWST) was launched on 25 December 2021. Equipped with its four instruments NIRISS\footnote{Near-InfraRed Imager and Slitless Spectrograph}, NIRCam\footnote{Near-InfraRed Camera}, NIRSpec\footnote{Near-InfraRed Spectrograph}, and MIRI\footnote{Mid-InfraRed Instrument} it has now started to provide its first observations in the infrared. Each instrument has different modes for photometry or spectroscopy and covers various regions of the spectra, from 0.6 to 28 $\mu$m. In particular, the instrument that covers the longer wavelengths is MIRI. 

The demands for observations with MIRI are high, and a significant part of the observations is dedicated to exoplanet observations, either directly or indirectly. When considering all Cycle 1 programs proposed for Early Release Science (ERS), Guaranteed Time Observation (GTO) and General Observations (GO), 115 distinct transiting exoplanets are being observed with the JWST. Twenty-one of these 115 planets have been observed with the MIRI instrument, 9 of them with the Low Resolution Spectrometer (LRS) \citep{kendrew_mid-infrared_2015}. These programs with MIRI LRS are mainly focused on small planets as this instrument is best-suited to observing the thermal emission of temperate rocky or sub-Neptune planets. The scientific impact of these observations is very high as the characterisation of rocky temperate exoplanetary atmospheres has just started \citep{greene_thermal_2023, zieba_no_2023}. In this context, the knowledge of the instrumental effects of MIRI LRS and its expected performance is key.

Although the quality of the first LRS data for transiting exoplanets is exquisite \citep{bouwman_spectroscopic_2022}, some instrumental effects remain poorly understood. In this regard, being able to create realistic simulated data that account for the specificity of the MIRI LRS data brings remarkable prospects in understanding these effects, in strengthening our data reduction methods, and in providing accurate spectra for characterising the physical and chemical composition of atmospheres. More than this, simulations are able to depict the accuracy of complex retrieval methods. The physical parameters that we expect to retrieve from the analysis of our simulations must be consistent with those that were injected. In addition, these simulations are very valuable for the community for confidently preparing upcoming observations, proving their feasibility, and therefore maximising their scientific outcomes. 

This paper is structured as follows. In Sect.~\ref{section:miri_lrs_simulations} we describe the simulation tool that we created and the detector modelling approach used in this work. In Sect.~\ref{section:case_l168} we use our tool on the science case of the transiting super-Earth L168-9b, an exoplanet that was observed as part of the MIRI LRS commissioning. Then, we compare our simulations with real data and show that we retrieve the same parameter values as we injected. In Sect.~\ref{section:nsr} we discuss how simulations can be used to identify unexpected behaviour in real data, and we provide a set of best practices to adopt for the reduction and analysis. Sect.~\ref{section:persistence-effect} focuses on persistence effects that have a major impact on the stability and therefore the accuracy of transiting exoplanet observations. Finally, Sect.~\ref{section:futur_dev} summarises our results and provides further routes for development.

\section{Simulations}
\label{section:miri_lrs_simulations}

To create time series of MIRI LRS spectra, we follow a four-stage process. First, we create the star-planet emission time series of 1D spectra with the \texttt{exoNoodle} package \citep{martin-lagarde_phase-curve_2020}. Then, we use \texttt{MIRISim} \citep{klaassen_mirisim_2020} to convert the astrophysical signal into detector spectral images by considering both the telescope optical path and the MIRI instrument transmission coefficients. At this stage, we also simulate the detector behaviour and add effects at the pixel scale, but over only one non-destructive integration. However, transiting-exoplanet simulations require another specific stage. The search for very faint flux variations requires considering faint detector persistence effects that may have an impact on the characterisation of the atmosphere. To include these features into the simulations, we created the \texttt{MIRISim-TSO} tool, which adds low-frequency detector persistence effects to time-series observations (TSO). When our simulations are complete, we proceeded with the data reduction steps.

\subsection{Star-planet system time series: \texttt{exoNoodle}}\label{section:exoNoodle}

The first step is to simulate an astronomical scene which is the incoming light from a star-planet system as the exoplanet orbits its host star. Light-curve simulation codes are quite numerous within the community. Mostly based on the \cite{mandel_analytic_2002} and the \cite{gimenez_equations_2006} analytic light-curve and limb-darkening models, some tools compute exoplanetary transit light curves, such as \texttt{PyTransit} \citep{parviainen_pytransit_2015}, the \texttt{TRIP} module of \texttt{ExoTETHyS} \citep{morello_exotethys_2020}, and even \texttt{PyPplusS}, which computes transiting exoplanets \citep{rein_fast_2019}. Other packages are designed to perform fits of exoplanet transits and radial velocity variations, including \texttt{TAP} \citep{gazak_transit_2012},  \texttt{EXOFAST} \citep{eastman_exofast_2013}, and  \texttt{JKTEBOP} \citep{popper_photometric_1981, southworth_eclipsing_2004}, which was created to fit light curves of eclipsing binary stars. More recently, new Python frameworks have been released and offer comprehensive models and fitting toolkits, such as \texttt{exoplanet} \citep{foreman-mackey_exoplanet_2021} and \texttt{starry} \citep{luger_starry_2019}. Both \texttt{batman} \citep{kreidberg_batman_2015} and \texttt{PyLightcurve} \citep{tsiaras_new_2016} provide exoplanet transit and occultation light curves, while \texttt{SPIDERMAN} \citep{louden_spiderman_2018} produces phase curves in addition to occultations. Light-curve observations made with the aim to characterise atmospheres are highly sensitive to the limb-darkening effect, and all codes pay particular attention to this. We mention here specific limb-darkening computation codes that can be added to the long list given above:  \texttt{ExoTIC-LD} \citep{laginja_exotic-ism_2020},  \texttt{exoCTK} \citep{bourque_exoplanet_2021},  \texttt{ExoTETHyS.SAIL} \citep{morello_exotethys_2020}, and  \texttt{Limbdark.jl} \citep{agol_analytic_2020}. 

Each one of these simulation codes brings a different perspective to the light-curve modelling: They either compute a transit, an occultation, or a phase curve based on the star-planet emission and atmospheric transmission. However, most of these codes simulate normalised light curves rather than absolute fluxes. In keeping with these models, \cite{martin-lagarde_phase-curve_2020} developed the \texttt{exoNoodle}\footnote{The code is available at \url{https://gitlab.com/mmartin-lagarde/exoNoodle-exoplanets}}. This is a Python tool that generates time series of spectra in absolute flux as the exoplanet orbits the star. The implemented model is based on the \cite{mandel_analytic_2002} prescription. Simulating spectra over the orbital phase allows us to prepare any MIRI LRS time-series observations in absolute flux.

To create these simulations, \texttt{exoNoodle} requires several inputs, which are listed below. 
\begin{enumerate}
    \item The emission spectrum of the star.
    \item The day- and nightside emission spectrum of the planet.
    \item The atmosphere transmission spectrum (i.e. in a transit geometry) in units of ${(R_{\mathrm{p}}/R_{\star})}^2$.
    \item The limb-darkening coefficients either as a quadratic or as a four-coefficient law. The limb-darkening coefficients may optionally depend on the wavelength.
    \item The orbital parameters: the orbital period $P$, the semi-major-axis $a$, the distance to the star $d$, the inclination of the orbit $i$, and the stellar and planetary masses and radii, $M_{\mathrm{p}},~R_{\mathrm{p}}$ and $M_{\star},~R_{\star}$, respectively.
    \item Phase values corresponding to the start and end of observation (between -1 and 1).
    \item The sampling time that corresponds to the interval between two spectrum computations in the time series. In the simulations presented in this paper, we decided to use the MIRI detector integration time. We calculated the integration time knowing the detector saturation level, the brightness of the target, and the LRS slitless frame time, taken from \cite{ressler_mid-infrared_2015}. We discuss the MIRI LRS readout pattern and terminology in more detail in Sect. \ref{section:mirisim}.
    \item A constant wavelength bin size $\Delta \lambda$ based on the MIRI LRS wavelength dispersion model from 5 to 12 $\mu$m and the spectral resolution model $R = \frac{\lambda}{\Delta \lambda}$ \citep{kendrew_mid-infrared_2015}. 
\end{enumerate}

As an output, \texttt{exoNoodle} creates a time series of spectra in $\mu \text{Jy}$. The number of spectra we compute is based on the duration of the time-series observation and the sampling we choose.

\subsection{Telescope and instrument simulations: \texttt{MIRISim}}
\label{section:mirisim}
        
After we modelled the time series of spectra with \texttt{exoNoodle}, we created spectral images of the observed scene with an instrument simulator. There are several tools that simulate detector images or spectra with a MIRI-like signal-to-noise ratio, such as the \texttt{PandExo} tool from \cite{batalha_pandexo_2017} and the Exposure Time Calculator for JWST \citep{pontoppidan_pandeia_2016}. To create realistic simulations of time-series observations, we use the stable version 2.4.2 of the MIRI official simulator \citep[\texttt{MIRISim}]{klaassen_mirisim_2020}. \texttt{MIRISim} simulates almost all MIRI observation modes (except for coronagraphy), including imaging, the LRS and Medium-Resolution Spectrometer (MRS) modes. The code itself is based on the use of calibration data products to mimic the telescope and instrument behaviour. The data products include the telescope optics diffraction, the filter transmission coefficients, and the detector dynamics. As input, this package takes the astronomical scene we produce with \texttt{exoNoodle} in $\mathrm{\mu}\mathrm{Jy}$. The input spectral absolute flux is then processed through the whole telescope path, including the instrument characteristics: the LRS subarray, the dispersion, and the detector photon-electron conversion efficiency. 

\subsubsection{The instrument model}
\label{section:mirisim_model}

Each spectrum is transformed into a spectral image as provided by the MIRI instrument LRS in slitless mode. The light coming from the point source is diffracted by the telescope optics and dispersed over the LRS slitless subarray (called SLITLESSPRISM), which is located at the top left of the MIRI focal plane \citep{ressler_mid-infrared_2015}. 
To simulate the dispersion of the diffraction pattern, a set of monochromatic normalised point-spread functions (PSFs), ${F}_{\mathrm{PSF_{norm}}}(\lambda)$, obtained from optical modelling and ground-based testing are positioned on the LRS slitless subarray following a wavelength-to-pixel dispersion law and a polynomial law of optical distortion. Then, the light coming from the point source is convolved with these normalised PSFs in order to obtain the PSFs in absolute flux. The light coming from the source ${F}_{\bigoplus\mathrm{ph}}(\lambda)$ and the absolute PSFs ${F}_{\mathrm{PSF}}(\lambda)$ are related through the $\mathcal{P}(\lambda)$ transfer function,
\begin{equation}
    {F}_{\mathrm{PSF}}(\lambda) = \mathcal{P}(\lambda)  \ast {F}_{\bigoplus\mathrm{ph}}(\lambda) \; ,
    \label{eq:psf_convolution}
\end{equation}
where $\mathcal{P}(\lambda)$ is given by
\begin{equation}
    \mathcal{P}(\lambda) = A \; {T}_{\mathrm{T}}(\lambda) \; {F}_{\mathrm{PSF_{norm}}}(\lambda)  \;,
    \label{eq:psf_transfer_function}
\end{equation}
${F}_{\bigoplus\mathrm{ph}}(\lambda)$ is the surface flux received by the telescope (in photon \si{\per\second\per\meter\squared\per\micro\meter}), subtended by the telescope entrance pupil area $A$ (in \si{\meter\squared}), and ${T}_{\mathrm{T}}$ is the dimensionless telescope transmission function. 

As the diffraction pattern is sampled into a given number of pixels, each pixel receives a sub-amount of the overall absolute PSFs flux. The absolute PSFs flux can be described as the sum of the flux that arrives at each pixel, 
\begin{equation}
    {F}_{\mathrm{PSF}}(\lambda) = \sum_{i=1}^{n} {F}_{\mathrm{pixel, i}}(\lambda) \; ,
    \label{eq:pixel_sampling}
\end{equation}
where $n$ is the number of pixels that form the diffraction pattern.

The pixel input flux ${F}_{\mathrm{pixel, n}}(\lambda)$ is then converted into an electronic signal ${S}_{\mathrm{pixel, n}}(\lambda)$, and both
quantities are related through the second transfer function $\mathcal{E}_{n}(\lambda)$,
\begin{equation}
    {S}_{\mathrm{pixel, n}}(\lambda) = \mathcal{E}_{n}(\lambda)  \times {F}_{\mathrm{pixel, n}}(\lambda) \; ,
    \label{eq:detector_product}
\end{equation}
where $\mathcal{E}_{n}(\lambda)$ is given by
\begin{equation}
    \mathcal{E}_{n}(\lambda) = \frac{\Delta \lambda \; {t}_\mathrm{frame} \times {tr}_{\mathrm{d}} \; \mathrm{QE}(\lambda)}{g}  \; ,
    \label{eq:detector_transfer_function}
\end{equation}
${S}_{\mathrm{pixel, n}}(\lambda)$ is obtained by integrating photons during a given interval of time ${t}_\mathrm{frame}$ (in \si{\second}), within a range of wavelengths $\Delta \lambda$ (in \si{\micro\meter}). The $\mathrm{QE}(\lambda) \; {tr}_{\mathrm{d}}(\lambda) $ product is called the photon-electron conversion efficiency (PCE; in $\mathrm{e}^{-} \: \mathrm{photon}^{-1}$) where $\mathrm{QE}(\lambda)$ is the detector quantum efficiency (in $\mathrm{e}^{-} \: \mathrm{photon}^{-1}$), and ${tr}_{\mathrm{d}}(\lambda)$ is the dimensionless transmission factor of the LRS double-prism assembly \citep{kendrew_mid-infrared_2015}. $g$ is the electronic gain that converts electrons into digital numbers (DN), in $\mathrm{e^-} \: \mathrm{DN}^{-1} $. The in-flight measurement of $\mathcal{E}_{n}(\lambda)$  during commissioning is called the absolute flux calibration factor and is expressed in 
($\mathrm{MJy} \: \mathrm{sr}^{-1}$)($\mathrm{DN} \: \mathrm{s}^{-1})^{-1}$ \citep{gordon_james_2022}. 

\begin{figure}
\centering
\includegraphics[width=0.5\textwidth]{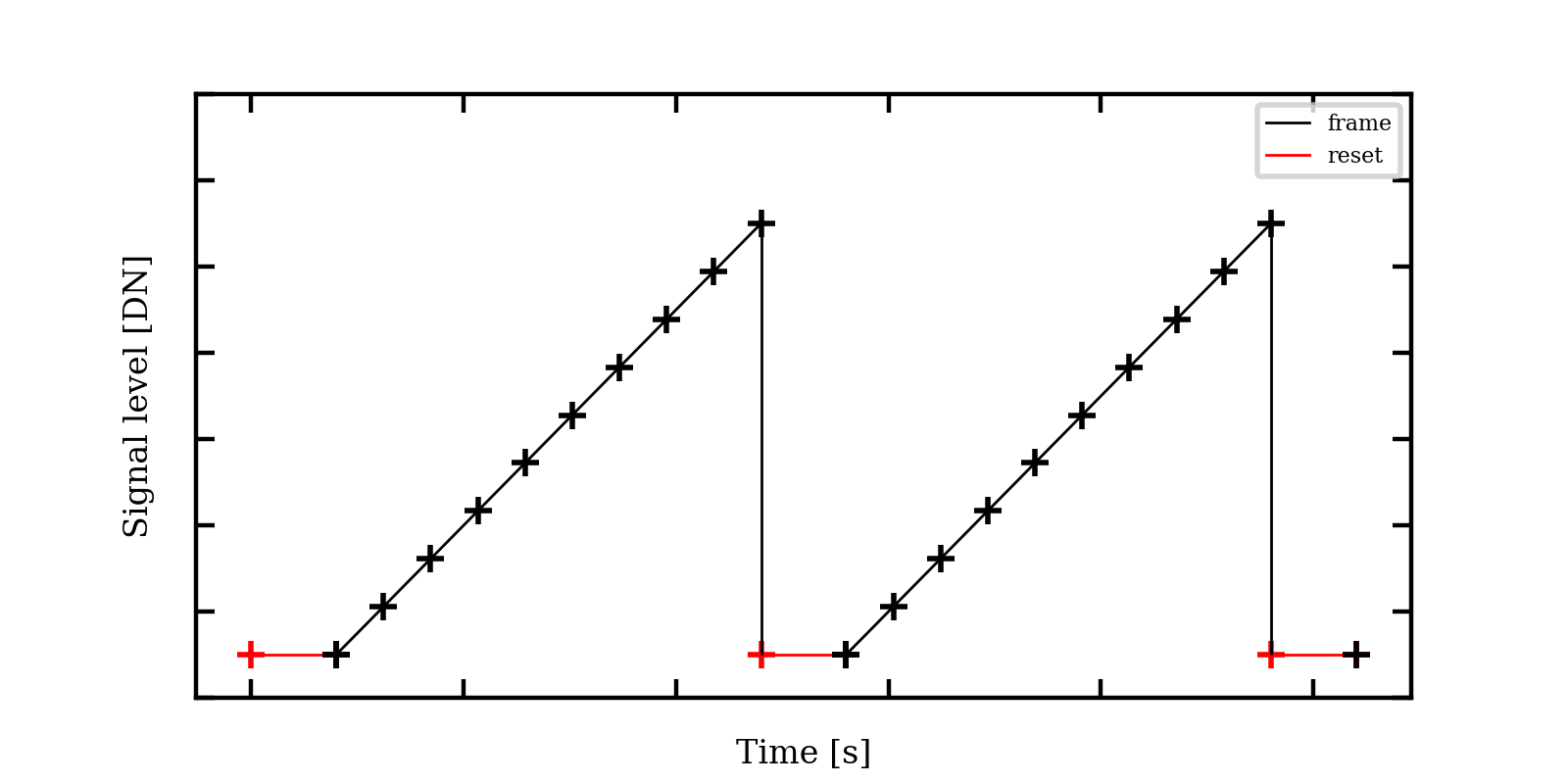}
\caption{Non-destructive readout pattern for the Mid-InfraRed Instrument Low Resolution Spectrometer (called FASTR1). Two consecutive integrations are displayed here. Each integration is composed of two resets and a series of frames. The number of frames is determined with the magnitude of the source and the detector pixel saturation level. An exposure is a set of several integrations. Adapted from \cite{ressler_mid-infrared_2015}.}
\label{fig:frame_miri}
\end{figure}

According to Eq. \ref{eq:detector_transfer_function}, photons falling into a pixel are converted into electrons and are then read out by the detector proximity electronics to be converted into DN. A readout pattern is a complete scheme of integrating light while doing multiple readings of a detector subarray. In the case of the MIRI detector, this scheme is called the FASTR1 non-destructive readout: The detector is read out non-destructively at regular intervals while integrating light, until it reaches a signal level close to saturation. After integrating light non-destructively, two resets are performed. This description is displayed in Fig.~\ref{fig:frame_miri}. The regular time interval between two readouts is called the frame time, which is 0.159s for the LRS slitless subarray \citep{kendrew_mid-infrared_2015}. This whole readout pattern composed of frames and two resets is called an integration, or a ramp. The integration time and therefore the total number of frames within an integration is chosen based on two parameters: the brightness of the observed target in $\mathrm{e^-} \: \mathrm{s}^{-1} $ and the detector pixel saturation level in $\mathrm{e^{-}}$. The whole observation is composed of several integrations and is called an exposure. 

Although resets are performed to empty the pixel potential wells, there is a remaining number of electrons that causes an offset at the beginning of each ramp. Originally, only one reset was performed within the readout pattern (formerly called the FAST mode). Then, to minimise detector systematics linked to the reset step, a second reset was added to the readout pattern. As a consequence, the offset value was lowered from 10 000 DN to 3000 DN \citep{argyriou_calibration_2021}. As \texttt{MIRISim} was coded based on the former FAST mode, we made the required changes in the \texttt{MIRISim} code to reproduce the FASTR1 mode. The offset value was changed to 3000 DN, and the timings were extended to include an additional reset. The whole reset pattern composed of two resets lasts 0.159 s, which is equivalent to a frame time.

\subsubsection{The instrument settings}
\label{section:mirisim_settings}

Simulations were made with the following \texttt{MIRISim} settings. As the observation mode is the LRS slitless mode, the focal plane subarray was set to SLITLESSPRISM. The filter parameter was set to P750L, which corresponds to the double-prism assembly. The telescope parameters were fixed to a beginning-of-life configuration, which is the telescope post-launch condition as determined during commissioning. The simulations included the following instrument settings: 
\begin{enumerate}
    \item A map of bad pixels (either hot or dead pixels) to be flagged as DO-NOT-USE in the simulations.
    \item A dark current map in $\mathrm{DN} \: \mathrm{s}^{-1} \:  \mathrm{pixel}^{-1} $. The dark current is a random generation of electrons through heat in the depletion layer, when no photons enter the detector. Decreasing the detector temperature is a way to limit the dark current \citep{glasse_mid-infrared_2015}.
    \item The flat-field map, that is, the relative response of pixels illuminated with a uniform source \citep{glasse_mid-infrared_2015}.
    \item The gain value in $\mathrm{e^-} \: \mathrm{DN}^{-1}$.
    \item Non-linearities affecting the ramp \citep{ressler_mid-infrared_2015, argyriou_calibration_2021}.
\end{enumerate}

Some effects that are newly witnessed in the data were not added to the simulations because no model has been released so far from either ground-based or in-flight testing. This is the case for the reset-switch charge decay (RSCD) \citep{ressler_reset_2023, morrison_jwst_2023}, caused by resetting the detector. This was not added in our simulations. The detector reset is accompanied by a transient effect that impacts the first frames of the subsequent integration, creating non-linearities at the start of the ramp \citep{rauscher_detectors_2007,  ressler_performance_2008, ressler_mid-infrared_2015}, which are only measured when the detector is in the dark (i.e. not illuminated). The reset anomaly is caused by trapped charges in the detector (possibly on the surface, at the indium bumps; see \cite{rauscher_detectors_2007}), even after reset and without illumination. When the detector is illuminated, the problems associated with the reset are known as the RSCD, which causes a greater increase in the ramp at the start of an integration (from the second integration onwards) and hence non-linearities. Finally, the reset causes a drop in offset at the start of the ramp, so that the starting point of the ramps would be different for the first and subsequent integrations. The addition of a second reset in the MIRIm ramps is partly linked to this problem and corrects offset errors at the start of the ramp after two resets \citep{argyriou_calibration_2021}. 

No background or noise are included at this stage of simulations. The readout noise in $\mathrm{e^{-}}$, caused by fluctuations in the readout amplifiers \citep{mcmurtry_james_2005}, is added to the simulations at this stage.

\subsubsection{The calibration data products}
\label{section:cdp}

\texttt{MIRISim} is based on the use of calibration data products (CDPs) that were initially created from ground-based testing campaigns. The in-flight calibration of the instrument made during commissioning provided a new set of files. These files are meant to be used by the Space Telescope Science Institute (STScI) \texttt{jwst} reduction pipeline\footnote{The \texttt{jwst} pipeline documentation is available at \url{https://jwst-pipeline.readthedocs.io/en/latest/getting_started/install.html}}. To create up-to-date simulations, CDPs were adapted to be compatible with \texttt{MIRISim}. They were taken from the \texttt{jwst} STScI pipeline calibration reference data system\footnote{The CRDS files are available at \url{https://jwst-crds.stsci.edu/}} (CRDS). CDPs of MIRI LRS taken from commissioning include non-linearity coefficients, dark map, readnoise map, bad-pixel mask, flat-field map, absolute flux calibration coefficients, spectral dispersion coefficients, gain value, and pixel area value.
Only the monochromatic PSF file comes from former optical modelling and ground-based calibration because no in-flight calibration file is available.
Commissioning revealed that the electronic gain value that converts electrons into DN is lower than expected. The simulations therefore include a gain of 3.1 $\mathrm{e^-} \: \mathrm{DN}^{-1} $  instead of 5.5 $\mathrm{e^-} \: \mathrm{DN}^{-1} $ (private communication, S. Kendrew,  2023)\footnote{In September 2023, the new gain values were ingested into the CRDS, \url{https://jwst-crds.stsci.edu/}, in context jwst\_1130.pmap}.

\subsection{Detector persistence effects: \texttt{MIRISim-TSO}}\label{section:MIRISIM_TSO}
     
In this section, we discuss the impact of detector persistence effects on the flux level over the whole exposure and how we include them in our simulations. Detector effects that were previously added in our simulations such as the dark current or the readout noise are additive and only depend on the pixel location on the detector. Non-linearities also affect the signal level, but only over one integration. In the end, \texttt{MIRISim} applies these effects independently on each integration. However, detectors feature low-frequency drifts that evolve over several integrations and might affect the whole exposure \citep{irac_instrument_and_instrument_support_teams_irac_2021}. These drifts account for 1 to 2 $\%$ of the absolute flux, and have the same order of magnitude or might even exceed the atmospheric feature amplitudes that we are meant to detect in our observations.
These temporal drifts may have different origins. An origin might either be a deviation in the temperature stability of the proximity electronics or a telescope-pointing deviation, known as jitter. Jitter is known to have a low impact on the MIRI detector outputs as commissioning confirmed a pointing stability better than expected of $\simeq$ 1 mas \citep{rigby_science_2023}. As the pointing stability is excellent and the PSF is well sampled from 5 to 12 $\mu \mathrm{m}$ and above, no intra-pixel noise is measured. This intra-pixel noise was known to be one of the most problematic issues for Spitzer data reduction \citep{ingalls_intra-pixel_2012, morello_repeatability_2016}. In the particular case of MIRI, detector persistence effects may also affect the flux stability. Some tests performed on the MIRI detector test model in 2018 by the Jet Propulsion Laboratory (JPL) that were presented in \cite{martin-lagarde_preparation_2020} show different types of persistence effects over an exposure. These data consist of photometric time series obtained by illuminating the detector with a set of blackbodies at different temperatures. The use of different blackbodies provided a set of light curves at different flux levels, expressed in $\mathrm{DN} \: \mathrm{s}^{-1}$. The independent fit of these light curves reveals that the persistence effects are flux-dependent. Using these test data, we identified and modelled these persistence effects and included them in our simulations. Analysis of the commissioning data indeed revealed these effects. Sect.~\ref{section:persistence-effect} broadly discusses their nature and the quantification of their impact on time-series observations.

\subsubsection{The \texttt{MIRISim-TSO} tool}
\label{section:mirisim_tso_tool}

Persistence effects are induced by previous uses of the detector. They therefore depend on its history. These previous operations tend to modify the detector behaviour at the beginning of an exposure. The direct consequence of these modifications is that the output flux level is altered and no longer corresponds to its expected value. The JPL tests of 2018 revealed three persistence effects in the data: the response drift, the anneal recovery, and the idle recovery \citep{martin-lagarde_preparation_2020}. Each one of these effects is described more extensively in Appendix~\ref{section:MIRISIM_TSO-description}. The \texttt{MIRISim-TSO} tool was created to add these effects to the simulations\footnote{The tool is available at \url{https://gitlab.com/mmartin-lagarde/mirisim_tso}}. 

All three persistence effects are described using an exponential model and come from \cite{martin-lagarde_preparation_2020}. They are implemented in this form in \texttt{MIRISim-TSO}. For all effects, the time variable $t$ varies between $t_{0}$ set to 0, the beginning of an observation, and $t_{0} + n_{\mathrm{int}}\Delta t_{\mathrm{int}}, n_{\mathrm{int}} \in \mathbb{N}$, where $n_{\mathrm{int}}$ is the number of integrations, and $\Delta t_{\mathrm{int}}$ is the integration time. 
The response drift effect was tested for fluxes within the $[0 \mathrm{DN}/\mathrm{s};5000 \mathrm{DN}/\mathrm{s}]$ range. It can be expressed as follows:
\begin{multline}
    {S}_{\mathrm{RD}}\big(t\big) = {S}_{0} + a_{1}({S}_{0}) \exp \Bigg( {- \frac{t}{\alpha_{1}({S}_{0})}} \Bigg) \\
    + a_{2}({S}_{0}) \exp \Bigg( {- \frac{t}{\alpha_{2}({S}_{0})}} \Bigg) \; ,
    \label{eq:drift}
\end{multline}
where ${S}_{0}$ is the expected flux level in $\mathrm{DN} \: \mathrm{s}^{-1}$, $a_{1}({S}_{0})$ and $a_{2}({S}_{0})$ are the amplitudes of the two exponentials in $\mathrm{DN} \: \mathrm{s}^{-1}$, and $\alpha_{1}({S}_{0})$ and $\alpha_{2}({S}_{0})$ are the time constants in seconds.

Anneal recovery has a similar aspect as the idle recovery. As soon as the annealing process is stopped and the detector is cooled down again, anneal recovery starts. This may be prior to the beginning of an observation. The time variable is therefore expressed as follows: $t+t_{\mathrm{A}}$, where $t_{\mathrm{A}}$ is a negative value that refers to the starting time of the anneal recovery. Anneal recovery is expressed as follows:
\begin{equation}
    {S}_{\mathrm{A}}\big(t\big) = {S}_{0} + b_{1} \exp \Bigg( {- \frac{t+t_{\mathrm{A}}}{\beta_{1}}} \Bigg) + b_{2}\exp \Bigg( {- \frac{t+t_{\mathrm{A}}}{\beta_{2}}} \Bigg) \; ,
    \label{eq:anneal}
\end{equation}
where ${S}_{0}$ is the expected flux level in $\mathrm{DN} \: \mathrm{s}^{-1}$, $b_{1}$ and $b_{2}$ are the amplitudes of the two exponentials in $\mathrm{DN} \: \mathrm{s}^{-1}$, and $\beta_{1}$ and $\beta_{2}$ are the time constants in seconds. 

The idle recovery effect is also expressed with one exponential. Its amplitude depends on the time spent resetting before the observation. Idle recovery is expressed as follows:
\begin{equation}
    {S}_{\mathrm{I}}\big(t\big) = {S}_{0}+ c({S}_{0},\Delta t_{\mathrm{I}} ) \exp \Bigg( {-\frac{t}{\gamma({S}_{0})}} \Bigg) \; ,
    \label{eq:idle}
\end{equation}
where ${S}_{0}$ is the expected flux level in $\mathrm{DN} \: \mathrm{s}^{-1}$, $c({S}_{0},\Delta t_{\mathrm{I}})$ is the amplitude of the exponential in $\mathrm{DN} \: \mathrm{s}^{-1}$, $\Delta t_{\mathrm{I}}$ is the time spent idling before the observation, and $\gamma({S}_{0})$ is the time constant in seconds. 

The input format of \texttt{MIRISim-TSO} is compatible with \texttt{MIRISim} outputs, that is, a series of integrations of 3D arrays in DN. The first two dimensions are the dimensions of the LRS slitless subarray (72 x 416 pixels), and the third dimension is the number of frames within one integration. To add persistence effects to the input data, Eq. \ref{eq:drift}, \ref{eq:anneal} and \ref{eq:idle} are integrated between $t$ and $t + \Delta t_{\mathrm{frame}}$, where $\Delta t_{\mathrm{frame}}$ is the frame time. Effects are therefore added frame per frame in DN. This operation is applied to each pixel independently. In this way, the flux dependence of each frame is taken into account at the pixel scale. 

\subsubsection{Adding the background to the simulations}
\label{section:background}

To include the background in the simulations, we used the background observation acquired on 26 May 2022 of the calibration target BD+60-1753. It consists of four LRS slitless integrations of 125 frames each. The flux level in $\mathrm{DN} \: \mathrm{s}^{-1}$ was then computed by fitting the slope over the integration time. To add the background in our simulations, we created a unique background image from all four integrations by taking the median value of the four images. Bad pixels were masked using the data-quality flags (DQ)\footnote{Data Quality flags are explained more extensively at \url{https://jwst-pipeline.readthedocs.io/en/latest/jwst/dq_init/index.html}}. These flags are meant to report any pixel issue that could be related to an unreliable behaviour of the detector. For example, flags report bad, hot, or saturated pixels. The slope values in $\mathrm{DN} \: \mathrm{s}^{-1}$ were integrated over the frame time to match the ramp structure of the data at a pixel scale. For each simulated target, the background image was scaled to match the observational background.

\subsubsection{Adding the photon noise to the simulations}
\label{section:photon_noise}

Photon noise was applied to the simulations in two steps. First, the signal S at the frame level in DN was converted into electrons using the electronic gain. Then, samples were drawn from a Poisson distribution and applied to the difference between pairs of frames. The number N of photons (or electrons in our case) received by a detector over a given time interval is described by the standard Poisson distribution,
\begin{equation}
    \operatorname{Pr}(N=k)=\frac{e^{-S g} \; (Sg)^k}{k !} \; ,
\label{eq:poisson_dist}
\end{equation}
where the $Sg$ product is the expected number of electrons, g being the electronic gain. 
The variance $\operatorname{Var}[N]$ of this distribution is
\begin{equation}
    \operatorname{Var}[N] = Sg \; .
\label{eq:variance}
\end{equation}

Photon noise in DN therefore is the standard deviation of the Poisson distribution and varies as the square root of the signal,
\begin{equation}
\sigma = \frac{\sqrt{S_{\mathrm{diff}} \; \mathrm{g} }}{\mathrm{g}} \; ,
\label{eq:photon_noise}
\end{equation}
where $S_{\mathrm{diff}}$ is the signal value in DN of the difference between pairs of frames at the pixel scale.

The output files of \texttt{MIRISim-TSO} were then combined into files of $2 \rm \; Go$, which correspond to the official segmented raw data products of JWST provided by STScI\footnote{The structure is the same as the uncalibrated \emph{\_uncal.fits} data files available on MAST \url{https://mast.stsci.edu/portal/Mashup/Clients/Mast/Portal.html}.}. These segmented files are 4D datasets with the same structure as the raw data products. The first two dimensions are the  dimensions of the LRS slitless subarray (72 x 416 pixels), the third dimension is the number of integrations, and the last dimension is the number of frames within one integration. In this way, the simulation output files are fully compatible with the STScI \texttt{jwst} reduction pipeline. 

\section{The case of L168-9b}
\label{section:case_l168}

In this section, we perform simulations of the transiting exoplanet L168-9b. JWST observed L168-9 as part of the MIRI LRS slitless commissioning under program ID 1033. The aim of this program was to test the time-series observation mode including the spectrophotometric stability. L168-9 is a bright M1V star located 25 pc away. It is orbited by a warm super-Earth that was initially announced by \cite{astudillo-defru_hot_2020}. L168-9b has a radius of $R_p = 1.39 \; \rm R_{\oplus}$ and a mass of $ M_p = 4.6 \; \rm M_{\oplus}$. This target was selected to meet the objectives of this calibration program because it probably shows no strong atmospheric features. The planet is expected to have an equilibrium temperature between 668 K and 965 K, and it has been found to be free from any primordial hydrogen-helium envelope \citep{astudillo-defru_hot_2020}. The outcomes of this program are presented in detail in \cite{bouwman_spectroscopic_2022}. 
L168-9b was observed on 29 May 2022 for $\simeq 4.2$ hours. The full observation is composed of 9371 integrations and represents $\simeq$ 6 Gb of data. 

We chose to reproduce the observation of L168-9b with the MIRI LRS slitless mode to fulfil two main purposes. First, our objective is to improve our simulations and make them as realistic as possible in order to prepare appropriately for future observation cycles. Then, our goal is to understand the origin of the systematics residuals that are witnessed in the data after reduction with the \texttt{jwst} pipeline \citep{bouwman_spectroscopic_2022}. The reduction pipelines are still under optimisation for time-series observations and play a part in the spectrophotometric precision that we obtain. Reliable synthetic data can therefore be used to improve the reduction methods. In this section, we present our simulations of L168-9b and their comparison to real data.

\subsection{Building the simulations}
\label{section:l168_simulations}

To be as realistic as possible, we used exactly the same setup for our simulations as in the real observations.
We simulated the observation of a transit of L168-9b following the approach detailed in Sect. \ref{section:miri_lrs_simulations}. 

\begin{figure}[ht!]
\resizebox{\hsize}{!}
        {\includegraphics[width=1.1\columnwidth]{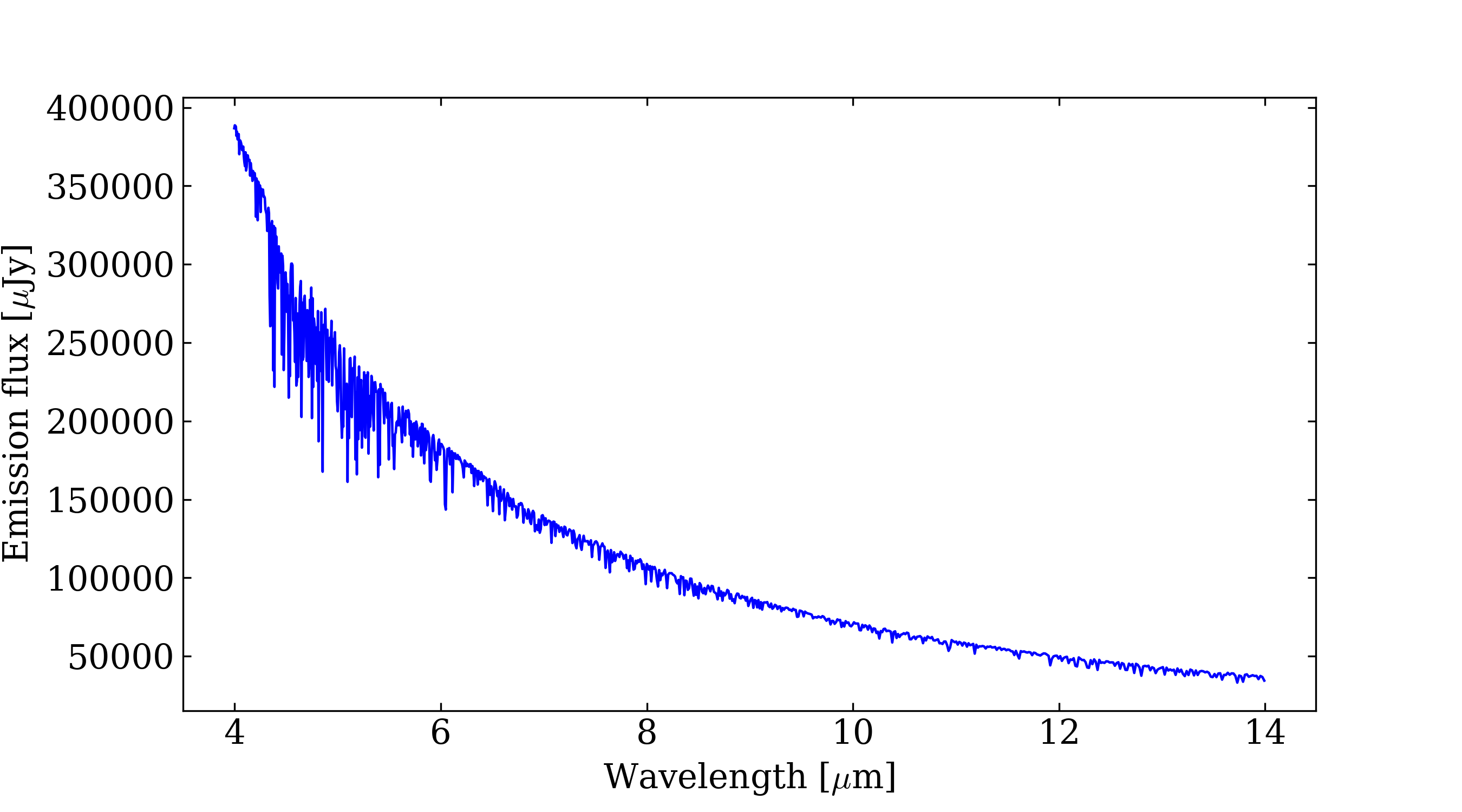}}
  \caption{L168-9 emission spectrum modelled with \texttt{PHOENIX} \citep{husser_new_2013}, interpolated to the appropriate temperature and metallicity.}
     \label{fig:L168-9_emission}
\end{figure}

As inputs for \texttt{exoNoodle}, we used a synthetic PHOENIX stellar spectrum \citep{husser_new_2013} interpolated to the appropriate temperature and metallicity. Fig.~\ref{fig:L168-9_emission} shows the stellar emission spectrum of L168-9. The planet parameters are derived from the literature \citep{astudillo-defru_hot_2020,patel_empirical_2022}, and Table~\ref{tab:input_param_simu} summarises all input parameters used for \texttt{exoNoodle}. 

\begin{table*}[ht!]
    \caption{Input parameters for \texttt{exoNoodle} to simulate a transit observation of L168-9b with the MIRI LRS slitless mode.}
    \label{tab:input_param_simu}
    \centering
    \begin{tabular}{c c c}
        \hline
        \hline
         Parameter & Value & Source \\
         \hline
         $T_{\mathrm{eff}}$ (K) & $3800 \pm 70$ & \cite{astudillo-defru_hot_2020} \\
         log $g$ (dex) & $4.04 \pm 0.49$ & \cite{astudillo-defru_hot_2020} \\
         metallicity [Fe/H] (dex) & $0.04 \pm 0.17$ & \cite{astudillo-defru_hot_2020} \\
         $R_{\mathrm{p}}/R_{\mathrm{star}}$ &  $0.0233 \pm 0.0007$ & \cite{patel_empirical_2022} \\
         ${M_\mathrm{p} \; (\rm M_{\mathrm{Jup}}}$) & $0.0145 \pm 0.0018$ & \cite{astudillo-defru_hot_2020} \\
         P (days) & $1.40150 \pm 0.00018$  & \cite{astudillo-defru_hot_2020} \\
         i ($\deg$) & $85.5 \pm 0.8$ & \cite{astudillo-defru_hot_2020} \\
         $a$ (AU) & $0.02091 \pm 0.00024$ &  \cite{astudillo-defru_hot_2020} \\
         \hline
    \end{tabular}
\end{table*}

\begin{figure}[ht!]
\resizebox{\hsize}{!}
    {\includegraphics[width=0.7\columnwidth]{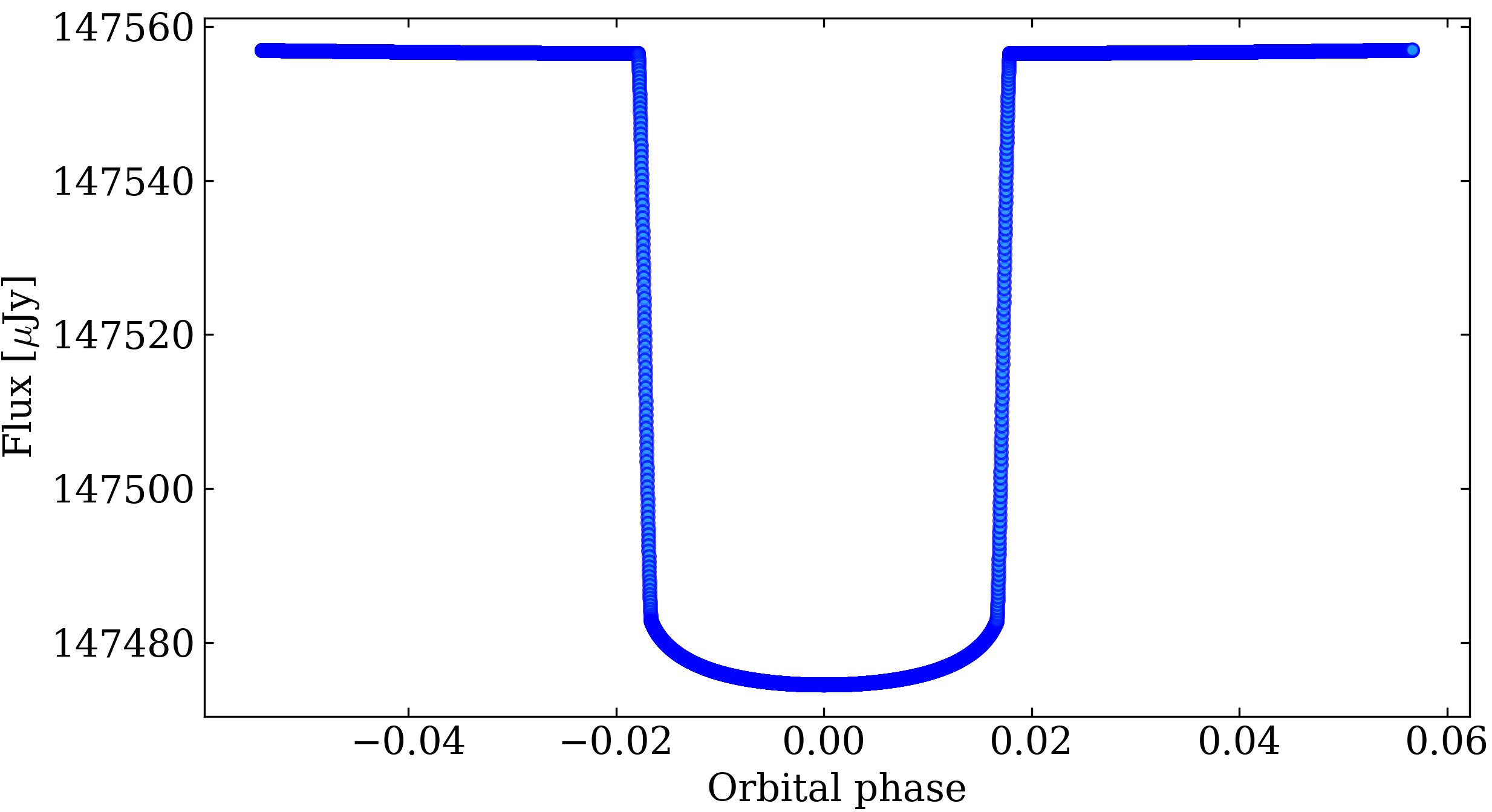}}
    \caption{L168-9b primary transit light curve in absolute flux, modelled with \texttt{exoNoodle}.}
    \label{fig:L168-9b_exo}
\end{figure}

For the sake of consistency with real data, the same integration time was used to sample the simulated time series with \texttt{exoNoodle}. Fig.~\ref{fig:L168-9b_exo} shows the associated white light-curve in absolute flux computed with \texttt{exoNoodle}. 
This white light-curve as well as the whole spectral time series was then provided to \texttt{MIRISim}. 
\texttt{MIRISim} was run using the LRS slitless mode, with nine frames and two resets per integration. The total number of integrations matching the data sampling on purpose, we obtained 9371 integrations. The \texttt{MIRISim} setup is detailed in Table~\ref{tab:mirisim_param_simu}. Using \texttt{MIRISim-TSO}, we added the background, the photon noise, and the persistence effects at the pixel and frame scale. 

\begin{figure}[ht!]
\resizebox{\hsize}{!}
    {\includegraphics[width=0.05\columnwidth]{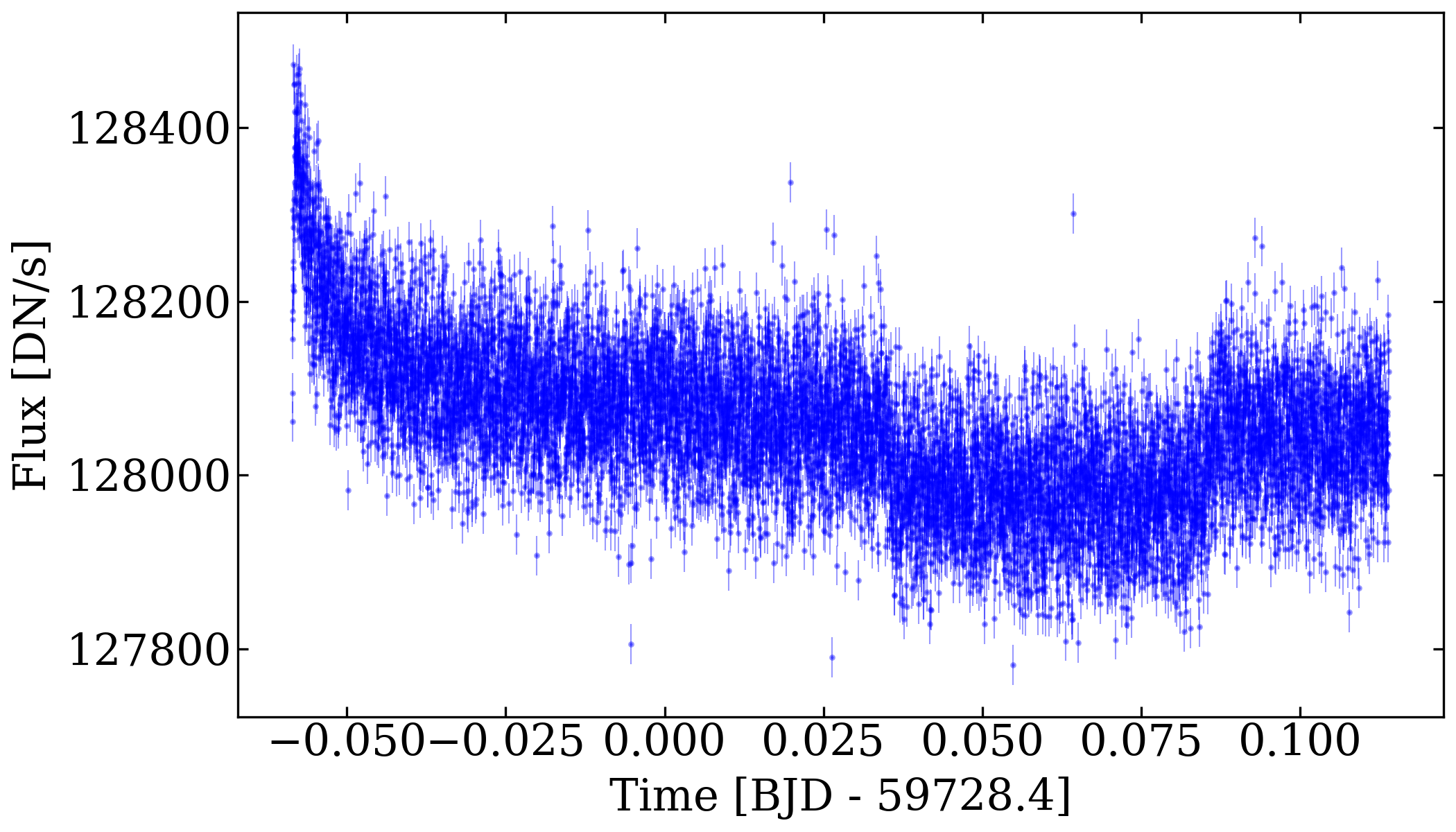}}
    \caption{L168-9b white light-curve extracted from the MIRI LRS data independently of \cite{bouwman_spectroscopic_2022}.}
    \label{fig:white_data}
\end{figure}

\begin{figure}[ht!]
\resizebox{\hsize}{!}
    {\includegraphics[width=0.05\columnwidth]{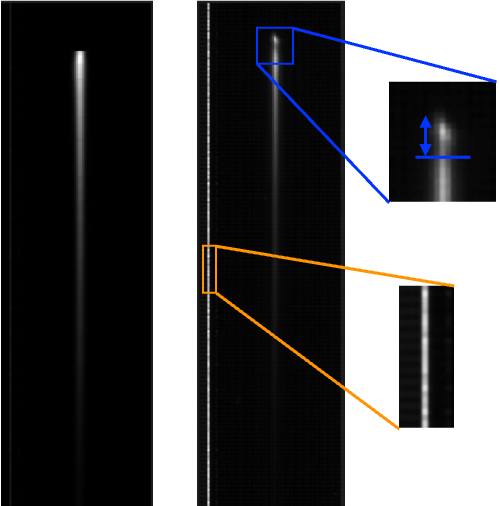}}
    \caption{MIRI LRS slitless subarray 416 x 72 pixels, cut at pixel 155 at the bottom of the spectra. L168-9b raw simulations and comparison to uncalibrated real data. The left panel shows the simulated spectral image, and the right panel is taken from the uncalibrated data. Both images are the last (ninth) frame of the first integration. The vertical and horizontal directions are pixels. The vertical axis is the spectral dispersion direction and is related to the wavelengths. Thus, pixels with high fluxes are located in the top part of the vertical axis, at short wavelengths. Then, the flux level decreases as the wavelength increases along the vertical axis. The real data display a hot pixel column on the left that is fully saturated (a small part of it is encircled in orange). The upper part of the trace shows two distinct zones. The top part, indicated with the blue arrow, shows a spectral contamination \citep{bouwman_spectroscopic_2022, bouchet_characterization_2022}.}
    \label{fig:uncal_compa}
\end{figure}

As mentioned in Sect.~\ref{section:MIRISIM_TSO}, we found evidence for persistence effects in the data that show an amplitude up to 1 to 2 \% of the absolute flux. Fig~\ref{fig:white_data} shows the overall aspect of the white light-curve of the time-series observation of L168-9b. 
We note a strong presence of an exponential decay in flux at the beginning of the observation. Based on the telemetry (see Appendix~\ref{section:telemetry}), several resets were performed prior to the observation, which means that the persistence effect visible in the white-light data is likely to be the idle recovery. 
To add the idle recovery to the simulations that comply with the real data, we fit the real white light-curve to derive the idle time parameter depicted in Eq.~\ref{eq:idle}. To do this, we selected a subarray centred on the spectral trace, which is two times the half-width source aperture (4 pixels). This corresponds to the pixels that are the most strongly impacted by the idle recovery. Then, we extracted more than 35 spectral light curves (following the process described in Sect.~\ref{section:reduction_3_5}). For each extracted light curve, we fitted a flux-independent Eq.~\ref{eq:idle}, and we derived the amplitude c and time constant $\gamma({S}_{0})$, knowing the $\Delta t_{\mathrm{I}}$ parameter, which is given by the telemetry in Fig.~\ref{fig:telemetry} of Appendix~\ref{section:telemetry}. Based on the amplitude and time-constant values obtained for the spectroscopic light curves, we linearly interpolated the values to create flux-dependent parameters that can be injected into Eq.~\ref{eq:idle}. Then we applied this idle formula to all pixels at different signal levels. We discuss the persistence effects witnessed in the real data in more detail in Sect.~\ref{section:persistence-effect}. As mentioned in Sect.~\ref{section:MIRISIM_TSO}, no jitter was added to the simulations because the stability of JWST is confirmed to be better than expected \citep{rigby_science_2023}. This hypothesis is indeed verified in the real data, for which no changes in position caused by the pointing stability are reported. The results of our simulations are displayed in Fig.~\ref{fig:uncal_compa}. The left panel shows a spectral image produced by the simulation, and the right panel shows its comparison to a real uncalibrated image. Even though the two are really similar, the real image displays some features that do not appear in the simulation. In the real data, we note a hot pixel column on the left that is fully saturated. The top part shows spectral contamination \citep{bouwman_spectroscopic_2022, bouchet_characterization_2022} that remains under investigation. The corresponding pixels are below the MIRI LRS dispersion range, between 4 and 5 $\mu$m. We chose not to replicate these features as they are systematically removed during the reduction steps.

\setlength{\tabcolsep}{7pt}
\renewcommand{\arraystretch}{1.2}
\begin{table}[ht!]
    \caption{L168-9b observation parameters used in \texttt{MIRISim}. All parameters come from the  observational setup of L168-9b validated by the Astronomer's proposal tool\protect\footnotemark.}
    \label{tab:mirisim_param_simu}
    \centering
    \begin{tabular}{c c}
    \hline
    \hline
         Parameter & Value \\
        \hline 
         Observation mode & MIRI LRS slitless \\
         Frame time (s) & 0.159 \\
         Number of frames per integration & 9 \\
         Integration time (s) & 1.431 \\
         Number of integrations &  9371 \\
         Number of exposures &  1 \\
         Number of resets & 2 \\
         \hline
    \end{tabular}
\end{table}
\footnotetext{The Astronomer's proposal tool can be found at \url{https://www.stsci.edu/scientific-community/software/astronomers-proposal-tool-apt}.}

\subsection{Data reduction and analysis}\label{data_red_analysis}

After the spectra were simulated, we proceeded to reduce them with the exact same methods as for the real data. In order to compare the results of our simulations, we also reprocessed the real data to verify our approach.
For the sake of consistency, we used the same tools and compared the outcomes at different stages of the data reduction and analysis. The two main tools are the \texttt{jwst} pipeline \citep{bushouse_jwst_2023}, version 1.8.5 under CRDS context 1075
for the data reduction steps and the \texttt{Eureka!} Python package \citep{bell_eureka_2022} for the analysis. This process was divided into five stages. The first two stages focus on detector-level corrections and calibration. The second stage includes background subtraction. Spectral extraction is performed in stage 3, and spectroscopic light curves are extracted in stage 4. The last stage is the light curve fitting with an astrophysical model and systematics detrending. 
Stages 1 and 2 were run with a slightly modified version of the default setup of the 1.8.5 version of the \texttt{jwst} pipeline, and stages 3 to 5 followed the steps defined in the \texttt{Eureka!} pipeline. In this section, we describe our approach for both the data reduction (stages 1 and 2) and the data analysis (stages 3 to 5).

\subsubsection{Stages 1 and 2}
\label{section:reduction_1_2}

Stages 1 and 2 from the \texttt{jwst} pipeline are meant to remove and correct for detector systematics. Stage 1 operates at the pixel and frame level to extract the mean count rate out of the non-destructive readouts of the detector. To do this, stage 1 is initialised using the DQ flags. Then, flagged pixels are used to generate a mask that is applied to the whole subarray and ensures that no unreliable pixels are included in further calculations. The next step is to apply the dark current correction by subtracting a 4D map of the dark current, interpolated towards the dimensions of the dataset.
Some effects do not yet benefit from corrections: the first frame effect, the last frame effect, the RSCD for at least the first four frames \citep{argyriou_calibration_2021} and saturated pixels at the end of the ramp. The first and last frame effects are a direct consequence of the detector reset and present parity effects. As the name suggests, these effects impact the first and last frames of an integration, which are systematically lower than the empirical values evaluated by linear ramp extrapolations. The value of the last frame is even lower for odd lines, whose reset voltage is influenced by that of the adjacent line \citep{ressler_mid-infrared_2015, argyriou_calibration_2021}. Whenever a pixel is impacted by these effects, it is just flagged as a DO-NOT-USE pixel. The first and last frames are systematically rejected instead of being corrected. After the correction steps and flagging steps are applied, cosmic rays are detected and replaced using the two-point difference method \citep{anderson_optimal_2011}. In the first version of the pipeline described here (version 1.5.3 and CRDS context 0916), the RSCD effect did not benefit from a correction or a dedicated step. In the next versions, the RSCD step was not activated because its impact is not known when going through the first reductions.

The main function of stage 1 is to fit the ramp in order to derive a mean count rate, which is the slope of the ramp. A least-squares minimisation method is applied to fit the ramp. The default ramp-fitting algorithm uses an optimal weighting of the ramp that gives an additional weight to the first and last frames of the ramp, based on the number of frames \citep{robberto_generalized_2013}. 

As non-linearities may distort the ramp, a non-linearity correction is applied before fitting the ramp. 
The output of stage 1 is a 3D time series of the spectral images. The resulting quantity is the mean count rate, which is a flux of DN that is a number of DN per second crossing the section of a pixel ($\mathrm{DN} \: \mathrm{s}^{-1} \: \mathrm{pixel}^{-1}$). This physical quantity can be directly related to a flux of photons after applying absolute flux calibration, which is the main goal of stage 2. 

Stage 2 uses as input a set of mean count rates in $\mathrm{DN} \: \mathrm{s}^{-1}$ at the pixel level and converts them into $\mathrm{MJy} \: \mathrm{sr}^{-1}$, applying a calibration factor determined during commissioning. In the peculiar case of transiting exoplanets, absolute flux calibration is skipped. As planetary flux variations are always relative to the stellar flux, the light curves are normalised, and therefore, no absolute calibration is required. Before calibration, each spectral image is divided by the flat-field reference image. The last step of stage 2 is to subtract the background from each spectral image. Because there is no slit to isolate the point source, slitless spectroscopy also integrates background flux over time, which can be removed. Background is subtracted following the method explained in \cite{bouwman_spectroscopic_2022}. Ten columns on the left and ten on the right side of the trace (column 36) are selected. Then, the median value is used as a reference to create a background image. This image is then subtracted from the spectral image. As each spectral image has its own background, this process has the benefit of removing any time-variable instrumental features from the data (such as the 390 Hz noise; private communication, M. Regan, 2023). Persistence effects are flux-dependent, and therefore, the background pixels corresponding to a low level of flux are less impacted by these effects. Removing the background does not allow us to correct for these effects in the pixels located in the spectral trace. We refer to Sect.~\ref{section:persistence-effect} for further explanations of persistence effects.
Then, a spatial filter of outlier detection is applied to remove any hot pixels that would have been left in the subarray. This technique consists of applying a running median on each column vertically to detect and reject any outlier (i.e. a hot pixel or a saturated pixel) left in the subarray based on a 5$\sigma$ rejection threshold. Finally, the spectrum is cut at pixel 395, around 4.5 $\mu$m to avoid scattered light at short wavelengths.

\begin{figure*}[ht!]
    \sidecaption
    \includegraphics[width=12cm]{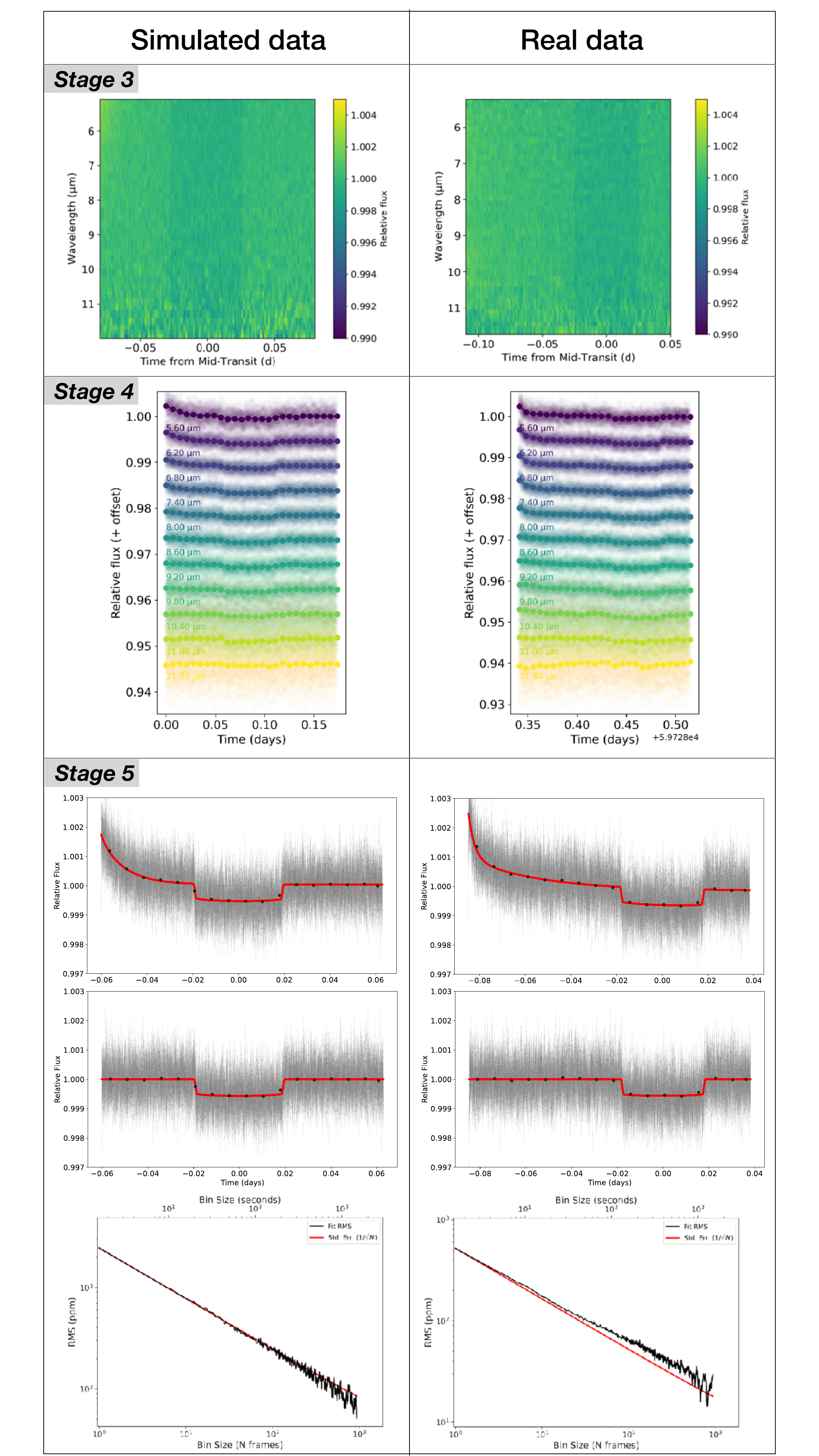}
    \caption{Comparison between real and simulated data at different stages of the reduction and analysis using \texttt{Eureka!} \citep{bell_eureka_2022}. {\it Top panel:} Waterfall plots from stage 3. {\it Centre panel:} Transit light curve as a function of wavelength, offset for clarity and produced with the \texttt{chromatic}\protect\footnotemark  visualisation tool. {\it Bottom panel:} Raw and detrended white light-curve with the best-fit model resulting from the optimisation performed in stage 5 (red curves) and the corresponding Allan plot, showing the evolution of the root-mean-square (RMS) of the light curve as a function of the binning size. The photon-noise limit is almost reached in both cases, attesting to the Gaussianity of the residuals. }
    \label{fig:eureka_compa}
\end{figure*}

The \texttt{jwst} pipeline relies on the use of reference and calibration files managed by the CRDS. To reduce our simulations and data, we used the latest in-flight version of these files. Only the electronic gain file was modified from 5.5 $\mathrm{e^-} \: \mathrm{DN}^{-1}$ to 3.1 $\mathrm{e^-} \: \mathrm{DN}^{-1}$ to comply with the value inferred during commissioning. The gain reference file has not yet been updated in the CRDS system, and the error arrays returned by the calibration pipeline therefore currently underestimate the true noise. 
\footnotetext{The \texttt{chromatic} package is available at \url{https://github.com/zkbt/chromatic}.}

\subsubsection{Stages 3 to 5}
\label{section:reduction_3_5}

For stages 3 to 5, we used the \texttt{Eureka!} pipeline \citep{bell_eureka_2022}. The input of stage 3 are \textit{.calints} files, which are spectral images produced by the \texttt{jwst} pipeline stage 2. In stage 3 the 1D spectrum is extracted from each image. Similarly to \cite{bouwman_spectroscopic_2022}, we chose a rectangular selection from pixel 155 to pixel 385 (height) and from pixel 13 to pixel 64 (width). We used a half-width source aperture of 4 pixels. The resulting waterfall plots from stage 3 for both real and simulated data are shown in the top panel of Fig~\ref{fig:eureka_compa}. The top panel shows the temporal evolution over the whole exposure represented on the y-axis in days, of each 1D spectrum displayed along the x-axis in $\mu$m. The colour bar refers to the level of normalised flux as a function of wavelength and time. The horizontal strip between 0.06 and 0.11 days in the left panel and that between 0.44 and 0.48 days in the right panel is darker than the rest of the time series. This deficit of flux corresponds to the planetary transit. The transit is well centred for simulations, but slightly offset in time for real data as some small uncertainties in the ephemeris of the planet existed when the commissioning observation were planned (in May 2022). In addition, the flux scatter is higher at larger wavelengths in general and at the beginning of the exposure at short wavelengths in particular. This is discussed in more detail in Sect.~\ref{section:nsr}.

\begin{figure}[ht!] 
    {\includegraphics[width=1.07\columnwidth]{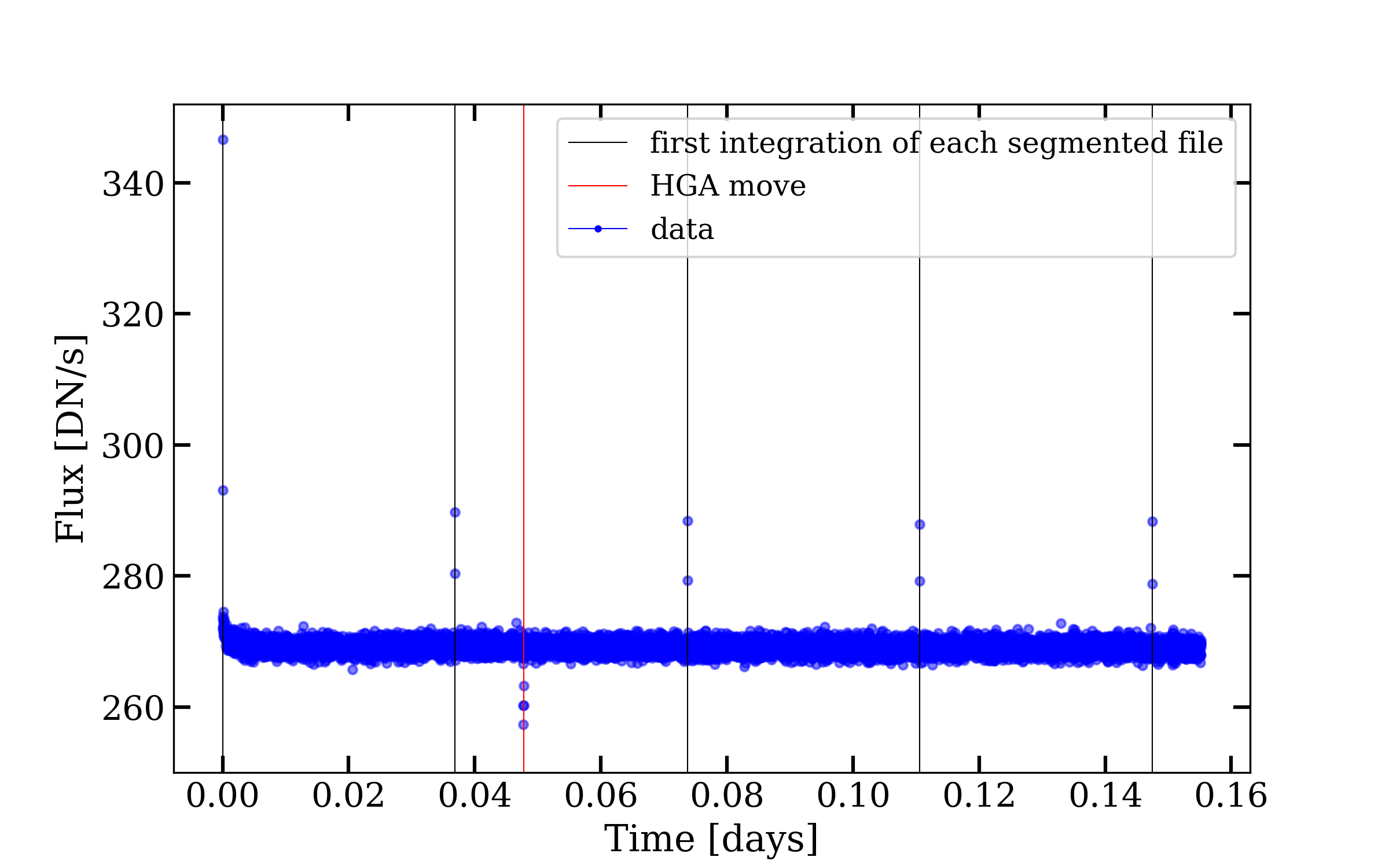}}
    \caption{White light-curve over the whole MIRI LRS slitless subarray of the real observation of L168-9b. Outliers are detected in the whole time-series data points. The black lines mark the position of the first integration of each segmented file, and the red line shows the timing of the High Gain Antenna (HGA) move during the exposure.}
    \label{fig:outliers}
\end{figure}

The main purpose of stage 4 is to produce a set of light curves that stem from spectral binning, which increases the spectral signal-to-noise ratio. To do this, we divided our dataset into 25 spectral bins, from 5 to 12.2 $\mu$m with a step of 0.145 $\mu$m. A subsample of 11 normalised light curves obtained from stage 4 of the 25 extracted bins is shown in the centre panel of Fig.~\ref{fig:eureka_compa} for the simulated and real data. Each light curve corresponds to the time series of a given spectral bin.
The second step of stage 4 is to apply a temporal sigma clipping of each light curve to remove the outliers. 
The \texttt{jwst} pipeline currently did not correctly handle dark current subtraction for segmented files. We therefore rejected the first integrations of each segment from the dataset. In the particular case of the L168-9b observations, some data points are also affected by the High Gain Antenna (HGA) move. Fig.~\ref{fig:outliers} shows the outliers witnessed in the real data. Each vertical black line marks the first integration of a segmented file, and the red line shows the HGA move during the exposure. We clipped the light curves using a box-car filter with a width of 100 integrations, with a maximum of ten iterations and a rejection threshold of 5$\sigma$ to reject these outliers.

Stage 5 fits both astrophysical and systematics models to infer the distributions of the transit parameters. Our fit was made using the no-U-turn sampler (NUTS) described in \cite{hoffman_no-u-turn_2011}, which is an extension of a Hamiltonian Monte Carlo algorithm that automatically tunes the step size and the number of steps per sample to avoid a random-walk behaviour. The transit was modelled using the \texttt{starry} package \citep{luger_starry_2019}. Instrumental systematics were modelled with a single exponential ramp. The jump parameters were the following:
\begin{enumerate}
    \item The radius ratio $R_{\mathrm{p}}/R_{\mathrm{\star}}$, assuming a wide uniform prior $\mathcal{U}(0.02 \pm 0.01)$ (the most recent value reported by \cite{patel_empirical_2022} is $R_{\mathrm{p}}/R_{\mathrm{\star}}=0.0233$).
    \item The reparametrised quadratic limb-darkening  coefficients ($q_{1}$, $q_{2}$) defined by \cite{kipping_observational_2016}, with uniform priors. This allowed them to vary between 0 and 1, $\mathcal{U}(0,1)$.
    \item The parameters of the one decaying exponential we used to correct for the persistence effect $r_0$ and $r_1$. The model of systematics is expressed as
    \begin{equation}
        S_{\mathrm{persistence}} = c_0 + r_0 \; e^{(-r_1 \; t)} \; ,
        \label{eq:exponentials_idle_S5_eureka}
    \end{equation}
    where $t$ is the time. We used loose normal priors derived from a preliminary least-squares fit on the white light-curve for $c_0, r_0, r_1$. These priors are reported in Table~\ref{tab:exponential_priors}.
\end{enumerate}

\setlength{\tabcolsep}{7pt}
\renewcommand{\arraystretch}{1.2}
\begin{table}[ht!]
    \caption{Priors used for the systematics detrending at stage 5 of \texttt{Eureka!}, derived from a preliminary least-squares optimisation.}
    \label{tab:exponential_priors}
    \centering
    \begin{tabular}{c c}
    \hline
    \hline
        Parameter & Prior distribution \\
        \hline 
        $c_0$ ($\mathrm{DN} \: \mathrm{s}^{-1}$) & $\mathcal{N}(1 ,0.1^{2})$ \\
        $r_0$ ($\mathrm{DN} \: \mathrm{s}^{-1}$) & $\mathcal{N}(-0.01 ,0.005^{2})$ \\
        $r_1$ ($\mathrm{s}^{-1}$) & $\mathcal{N}(75 ,30^{2})$ \\
        \hline
    \end{tabular}
\end{table}

The following parameters were first fitted on the white light-curve and then fixed for the analysis per wavelength to avoid variations that would be non-physical.
\begin{enumerate}
    \item The time of mid-transit $t_0$, with a normal prior of $\mathcal{N}(59728.4603 ,0.005^{2})$ $\mathrm{BMJD}_{\mathrm{TDB}}$ computed for the nearest epoch based on the best-fit ephemeris derived from the analysis of the most recent TESS observations (courtesy of Billy Edwards). 
    \item The orbital period $\mathrm{P}$, with a normal prior based on the value reported by \cite{astudillo-defru_hot_2020} $\mathcal{N}(1.40150 ,0.00018^{2})$. 
    \item The inclination $i$, with a normal prior based on the value reported by \cite{astudillo-defru_hot_2020} $\mathcal{N}(85.5 , 0.7^{2})$
    \item The eccentricity $e$, with a normal prior $\mathcal{N}(0. , 0.1^{2})$ taken arbitrarily from the unique known constraint, that is, $e<0.21$ \citep{astudillo-defru_hot_2020}.
\end{enumerate}

\begin{figure}[ht!]
    \centering
    \includegraphics[width=1.\columnwidth]{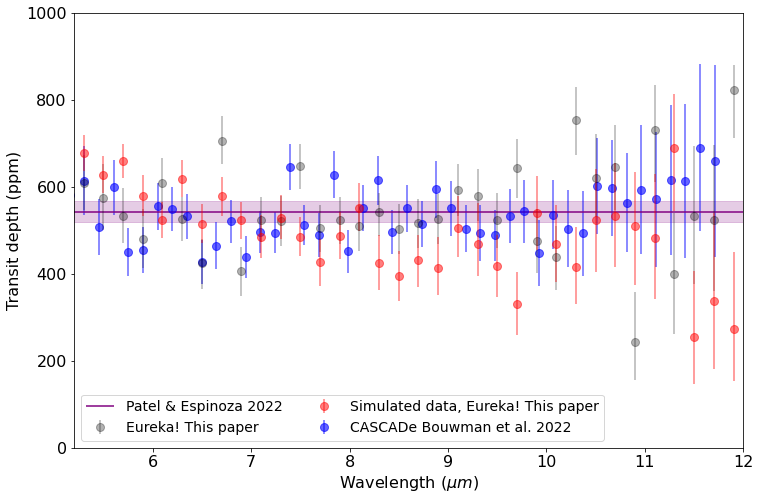}
    \caption{Transmission spectrum of L168-9b obtained from simulated (red dots) and real data (grey dots). The blue dots are the points from \cite{bouwman_spectroscopic_2022}, and the purple curve is the reference value from \cite{patel_empirical_2022} along with the 95\% confidence interval.}
    \label{fig:spectra_comparison}
\end{figure}

Using the probabilistic programming framework \texttt{pymc3}, which  features a \texttt{NUTS} sampler, we ran three chains with a number of iterations to tune the set to 2000 and a number of draws set to 2000. In the bottom panel of Fig.~\ref{fig:eureka_compa}, we show the resulting best fit of the white light-curve from real and simulated data using \texttt{NUTS}. The root-mean-square (RMS) of the detrended white light-curve is 501 ppm and 519 ppm for simulated and real data, respectively.
Table~\ref{tab:param_comparison} shows the comparison between the input parameters and the retrieved parameters for the simulations. Finally, stage 6 produces a transmission spectrum for the simulated and real data. Fig.~\ref{fig:spectra_comparison} shows the transmission spectra obtained from the analysis of the simulated and real data. We demonstrate that the simulations replicate the data with similar scatter and error bars, thus leading to a mean transit depth of $569 \pm 92$ ppm for the real data and $531 \pm 90$ ppm for the simulated data. 

\begin{table}[ht!]
    \caption{Retrieved parameters from the white light-curve fit and comparison to the \texttt{exoNoodle} input parameters}
    \label{tab:param_comparison}
    \centering
    \begin{tabular}{c c c}
        \hline
        \hline
        Parameter & Input value & Retrieved value \\
        \hline
        $R_{\mathrm{p}}/R_{\mathrm{star}}$ & $0.0233 \pm 0.0007$ & $0.023292^{+0.00028}_{-0.00028}$ \\
        $a/R_{\mathrm{star}}$ &  $7.493^{+0.610}_{-1.640}$  &  $7.443^{+0.840}_{-0.820}$ \\
        q1 & $0.0744906$ &  $0.18^{+0.40}_{-0.16}$ \\
        q2 & $0.09056845$ & $0.40^{+0.37}_{-0.26}$ \\
        \hline

    \end{tabular}
\end{table}

To confirm our result, we performed a parallel analysis with two other sampling methods: \texttt{emcee} \citep{foreman-mackey_emcee_2013}, and \texttt{dynesty} \citep{speagle_dynesty_2020}. We obtained very consistent results, meaning that the transmission spectrum is not dependent on the sampling method.

\subsection{Resulting transmission spectra and noise-to-signal ratio}\label{section_3_results}

\begin{figure}[ht!]
    \centering
    \includegraphics[width=1.\columnwidth]{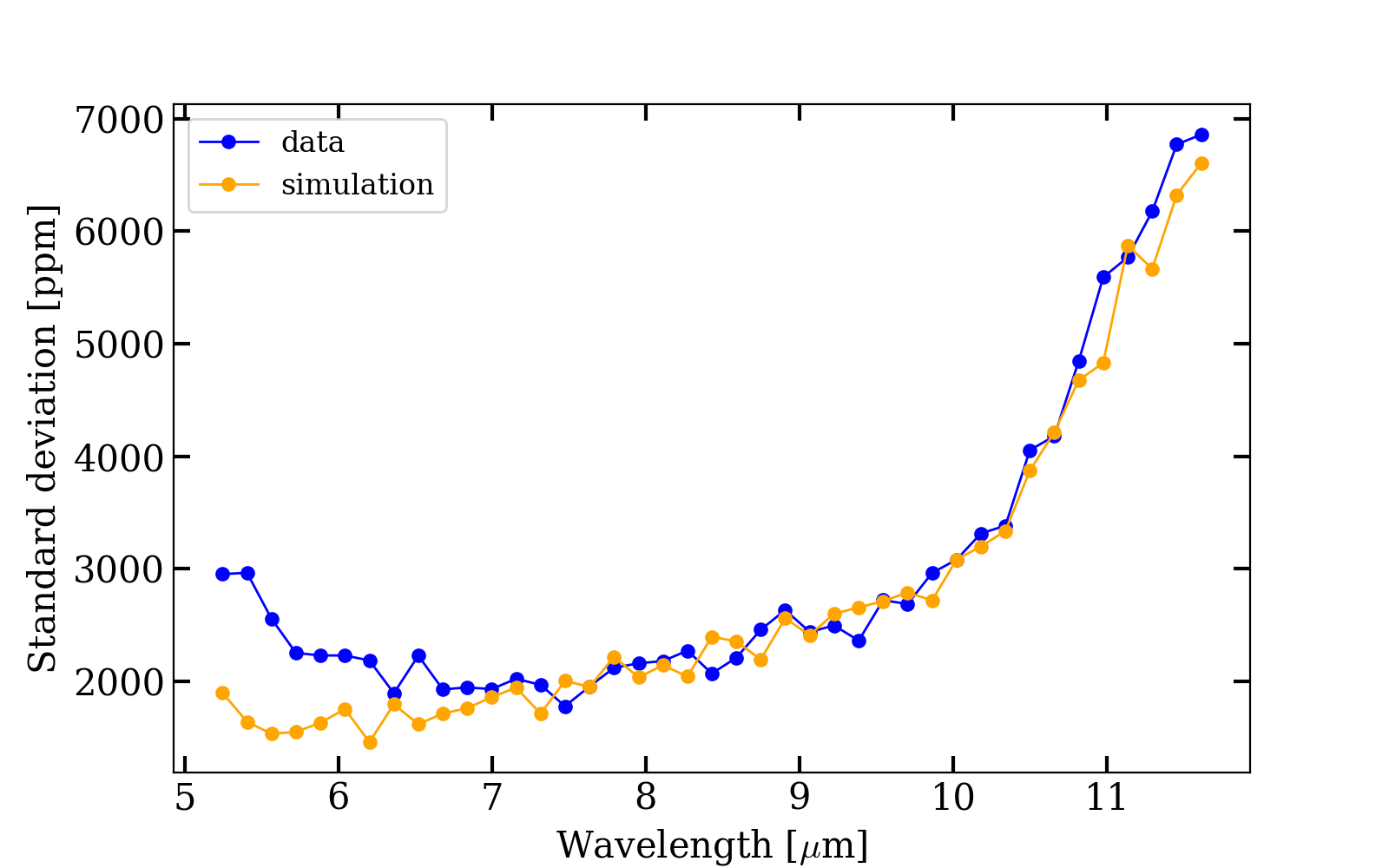}
    \caption{Noise-over-signal ($N/S$) estimates of the spectral light curves of L168-9b for the simulations and real data. The blue curve shows the estimate for the $N/S$ of the data, and the orange curve shows the estimate for the $N/S$ of the simulation.}
    \label{fig:error_vs_lambda}
\end{figure}

Fig.~\ref{fig:error_vs_lambda} shows the evolution of the noise-over-signal ($N/S$) estimate of the spectral light curves for simulations and real data. The $N/S$ was calculated based on the standard deviation of an out-of-transit part of the normalised light curves (between integrations 3500 and 4000). This $N/S$ computation method is equivalent to the method used in \cite{bouwman_spectroscopic_2022}, which plots the error bars obtained after the spectral fits because there is no correlated red noise in the residuals. The Allan plots of the residuals for spectral bins between 5 and 7 \si{\micro\meter} are shown in Appendix~\ref{section:correlated_noise}. 
In Fig.~\ref{fig:error_vs_lambda}, the simulation mimics the real data very well, except at short wavelengths, where the real data show an excess of $N/S$ up to 30\% between 5 and 7 $\mu$m. We do not expect a difference like this between the two datasets. This same behaviour of the real data noise estimate is seen in \cite{bouwman_spectroscopic_2022}. This additional $N/S$ at short wavelengths does not have any known physical interpretation, but it results in larger error bars for the transmission spectrum at shorter wavelengths and may consequently weaken future atmospheric retrievals. In Sect.~\ref{section:nsr}, we investigate this additional $N/S$ and provide a correction for it.

\section{The noise-to-signal ratio}
\label{section:nsr}

In this section, we investigate the additional $N/S$ based on the differences we noted between real data and simulations in order to correct for it. We first focused on targeting the origin of this additional $N/S$ to understand whether it is intrinsically present in the uncalibrated data or is linked to the data reduction methods. 

\subsection{An alternative data reduction}
\label{section:no_ramp_fitting}
To understand whether this excess $N/S$ measured at short wavelengths is an intrinsic feature of the data, we used an independent reduction method and compared our results to those obtained with the \texttt{jwst} pipeline (version 1.8.5, CRDS context 1075). This reduction method is intentionally simpler than the method offered by the pipeline. It is based on the use of the penultimate frame value of the ramp alone, instead of fitting the whole ramp, as performed in stage 1. 
The method of this alternative reduction and the results are given in Appendix~\ref{section:no_ramp_fit_method}. The standard deviation computation shows that there is no additional $N/S$ at short wavelengths in the L168-9b data after using this alternative data reduction. This is shown in Fig.~\ref{fig:NSR_new_method}. This confirms that the additional $N/S$ does not come from the data itself, neither from the source nor from the instrument, but from the data reduction method that is applied.

\subsection{Investigating the \texttt{jwst} pipeline steps}
\label{section:ramp_fitting}

\begin{figure*}[ht!]
    \sidecaption
    \includegraphics[width=12cm]{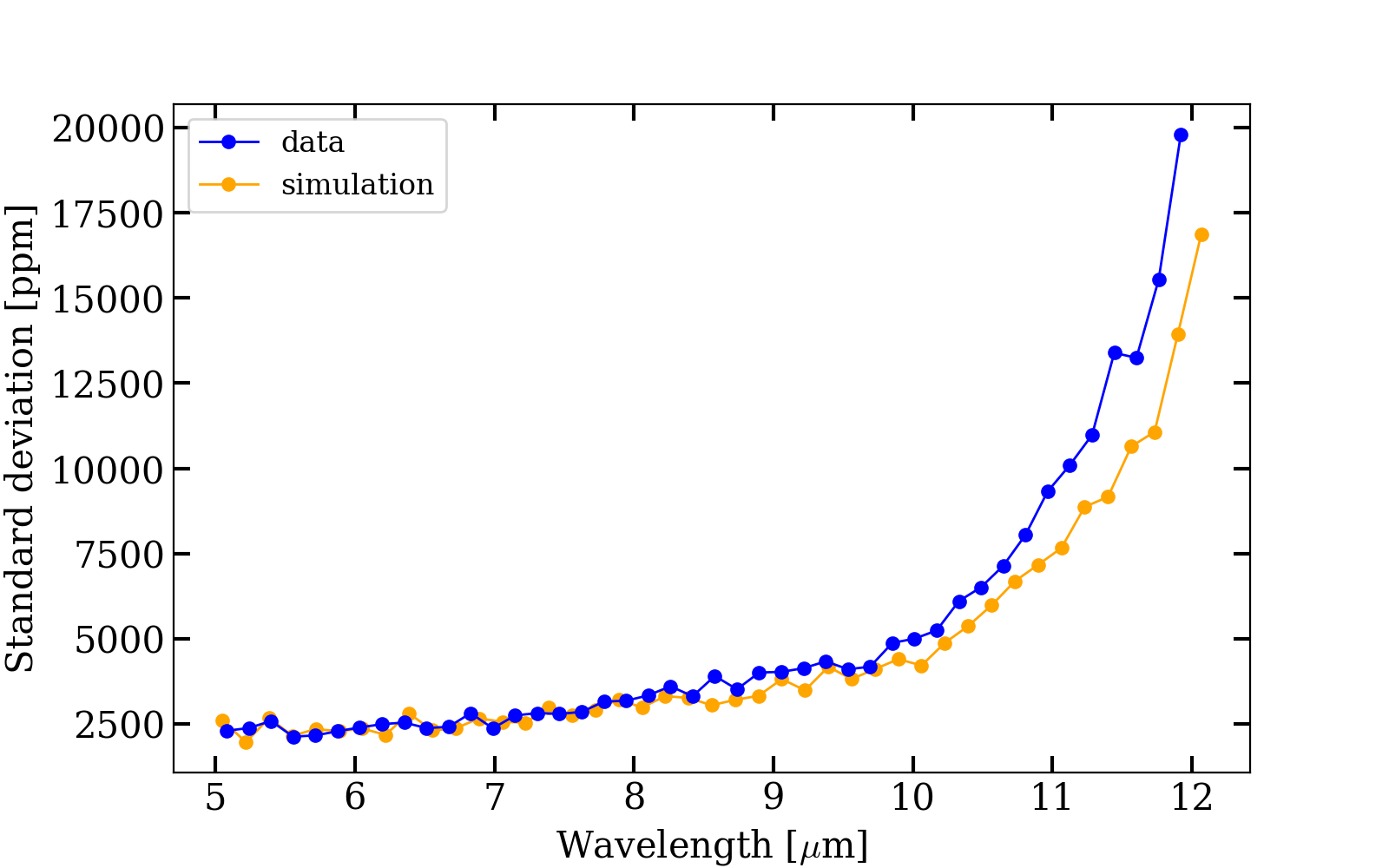}
    \caption{Noise-over-signal ($N/S$) estimates of the L168-9b observation and simulation after removing the first four frames of each integration of the exposure before fitting the ramp using the \texttt{jwst} pipeline Stage 1. No additional $N/S$ at short wavelengths is witnessed any longer.}
    \label{fig:nsr_rscd}
\end{figure*}

To conduct our analysis, we focused on stage 1 of the pipeline and isolated each step to test its impact on the output $N/S$. In this way, we alternately included all the steps before fitting the ramp: dark subtraction, corrections of saturation, reset anomaly, first-frame effect, last-frame effect, non-linearities, RSCD, cosmic rays, and finally, gain scaling. Among these steps, the last-frame effect, the first frame effect and the RSCD were not corrected but the impacted frames were flagged and excluded from any other step. This work shows that after activating the RSCD step, the L168-9b data $N/S$ no longer shows an increase at short wavelengths. Fig.~\ref{fig:nsr_rscd} shows the $N/S$ estimate for the observations and simulations. No excess in the $N/S$ between 5 and 7 \si{\micro\meter} is visible any longer.

\begin{figure*}[ht!]
    \sidecaption
    \includegraphics[width=12cm]{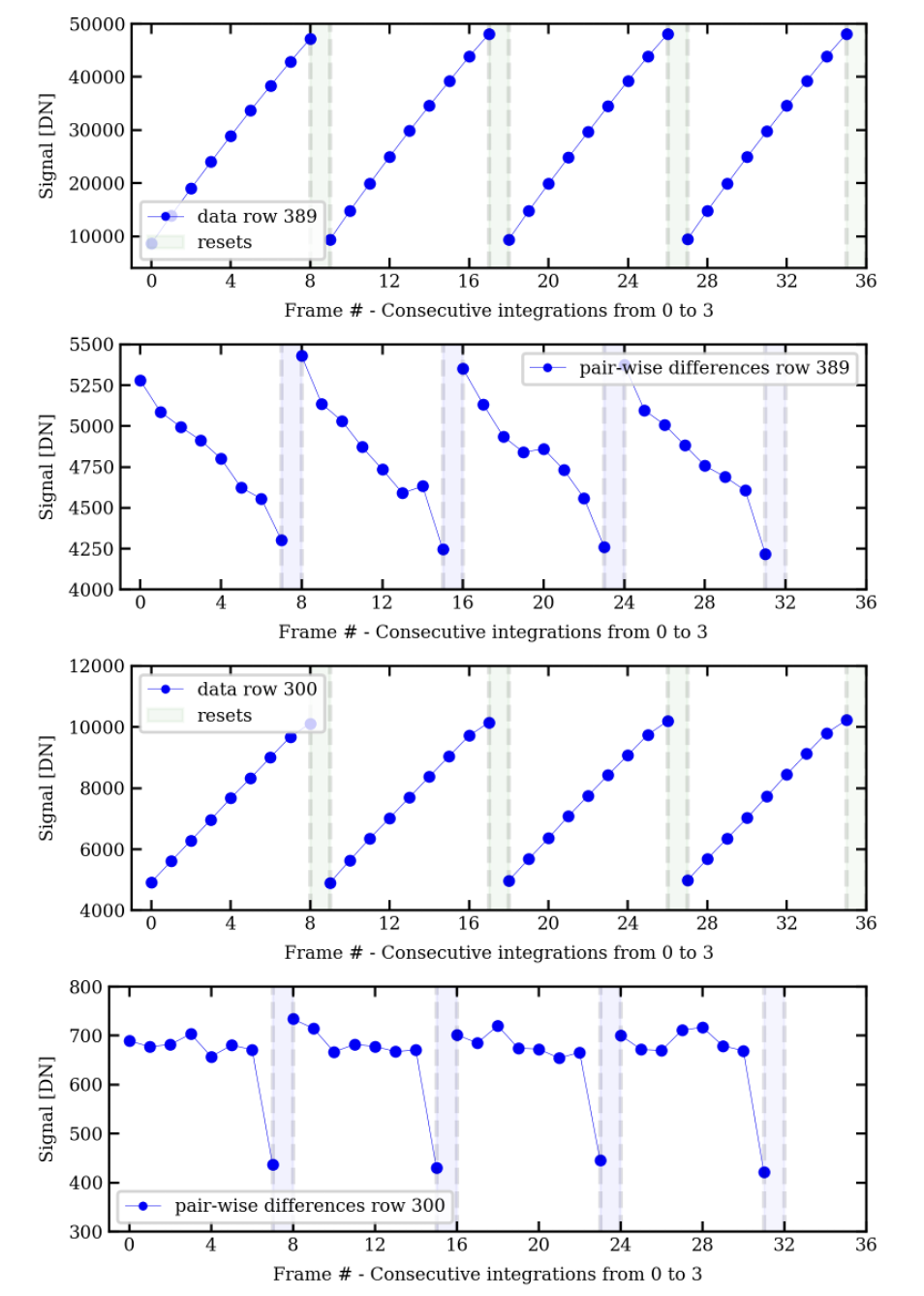}
    \caption{First four integrations of L168-9b target data after application of the non-linearity correction step. Figures 1 and 3 (starting from the top) show the ramps of pixels [389, 36] and [300, 36], where 36 is the brightest column. The two-pixel rows (389 and 300) are representative of a strong and medium signal, respectively. Figures 2 and 4 show the successive frame differences, where non-linearities still remain even after the correction.}
    \label{fig:ramps_after_correction}
\end{figure*}

The reason for this increase in $N/S$ at short wavelengths is complex. To provide a clear explanation of it, we start by recalling the cosmic-ray detection process. This process is based on the two-point difference method \citep{anderson_optimal_2011}, which rejects outlier frames that deviate from the mean value of the successive frame differences. When no cosmic ray or other non-linearity occurs in the ramp, the shape of the successive differences is expected to be a horizontal straight line. In reality, the ramps display residual non-linearities that remain at the end of stage 1, just before fitting the ramps. Fig.~\ref{fig:ramps_after_correction} shows the first four integrations of the data after correcting from ramp non-linearities in Stage 1. The figure shows the ramps of the brightest pixel [389, 36] (top panel) and those of a pixel with a medium signal [300, 36] (third panel from the top). Panels 2 and 4 show the successive frame differences of these first four integrations. We observe non-linearity residuals of up to 4\% for high signals. This value meets the 5\% requirement made during ground-based testing, but is still not sufficient.

\begin{figure*}[ht!]
    \sidecaption
    \includegraphics[width=12cm]{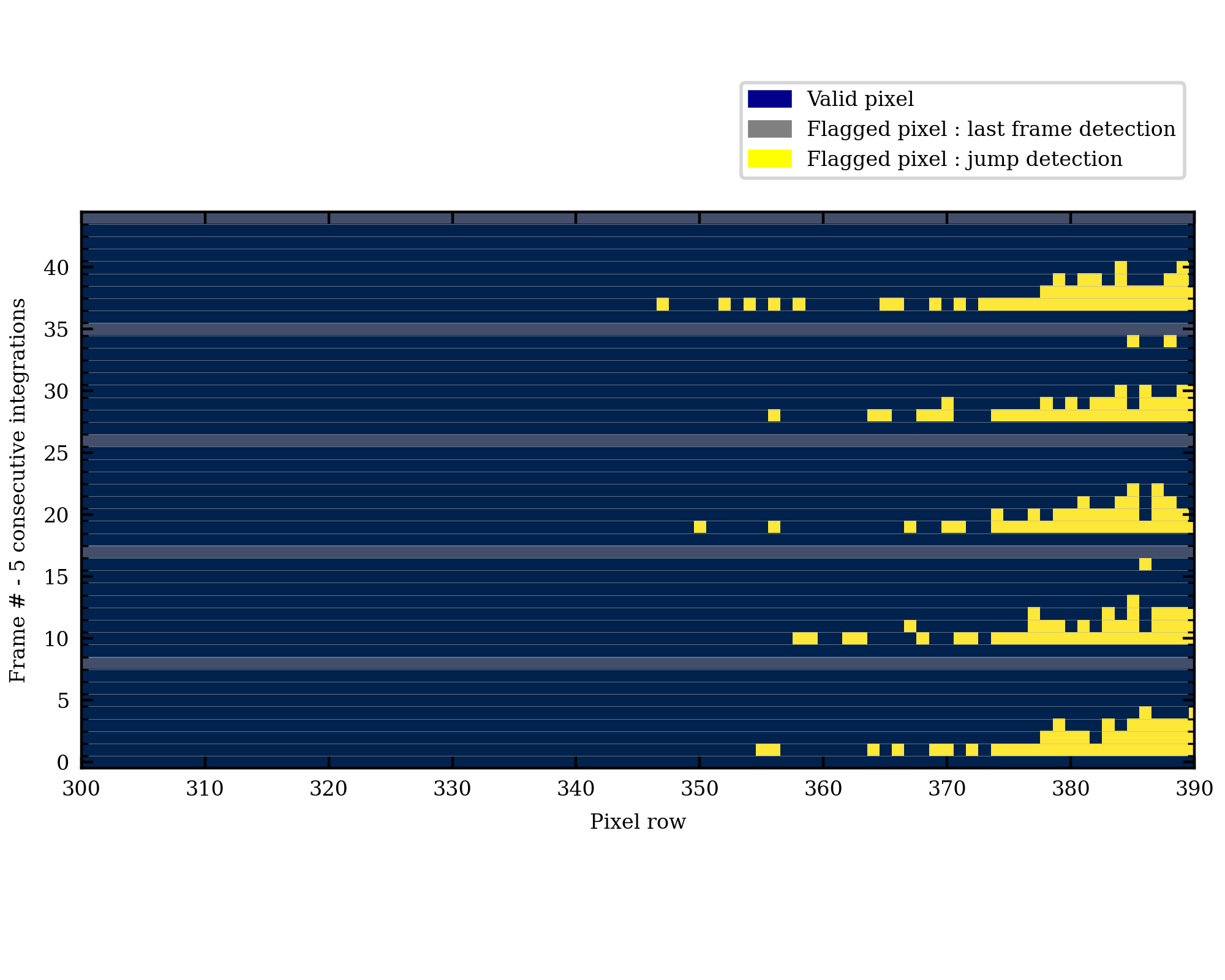}
    \caption{Flagging status at the frame level for the first five integrations (y-axis) of the L168-9b data after applying the cosmic-ray detection step (also known as the jump detection step) using version 1.8.5 of the pipeline. The flagging status is shown for all pixels located in the brightest column, 36 (x-axis). The colour code shows the status of each frame. Frames in blue are valid and therefore kept in the dataset for the ramp adjustment step. Frames in grey are rejected from the dataset by the last frame-effect detection step. Frames in yellow are rejected by the jump detection step and are mostly between number 1 and 5. They are located at short wavelengths (rows of pixels ranging from $\sim$ 340 to 390, mainly focused on pixels with high fluxes).}
    \label{fig:flags}
\end{figure*}

Because of these residual non-linearities, the cosmic-ray rejection stage flags frames that are not impacted by cosmic rays, but by these non-linearity residuals. These particular frames are mostly between frames 1 and 5. However, the first frame being affected by the first frame effect is never excluded from the dataset because it is not taken into account in the two-point difference method. Fig.~\ref{fig:flags} shows the flagging status of all nine frames for the first five integrations for all pixels of column 36 (rows ranging from 300 to 390).

Fig.~\ref{fig:flags} displays a fluctuation in the number of frames flagged by the cosmic-ray detection step for different integrations (depicted in yellow), which creates a variability in the ramp adjustments as the number of frames considered for the adjustment varies significantly only at high fluxes, that is, at short wavelengths. The resultant slope values are therefore highly scattered at short wavelengths, thus creating the excess of $N/S$ observed in the data. At medium and low flux level, we do not observe any additional frame rejection by the cosmic-ray detection stage.

The residual non-linearities may indeed be due to the RSCD effect or to any other detector effect \citep{argyriou_calibration_2021, argyriou_brighter-fatter_2023}. As mentioned in Sect.~\ref{section:mirisim_settings}, RSCD is caused by resetting the detector \citep{ressler_reset_2023} and generates non-linearities at the beginning of the ramp, which appear starting from the second integration. After commissioning, \cite{morrison_jwst_2023} investigated this effect on dark exposures and concluded that the RSCD displays two features. The fast decay appears in the second integration, where an additional signal at the beginning of the ramp decays exponentially. The first integration does not exhibit this decay as the preceding detector idling prevents any signal from accumulating in the detector traps. The second decay is a slope difference between integrations and depends on the number of frames. It appears that shorter integrations display larger differences that become even larger for the LRS subarray. 
The RSCD effect does not benefit from any correction in the pipeline, but the corresponding RSCD step flags the impacted four first frames. In other words, by activating the RSCD step, we remove the first four frames from the dataset.
When the RSCD step is activated, which is applied upstream of the cosmic-ray detection step, the first four frames of the dataset are systematically excluded from the ramp-fitting step, thus limiting the variability of the number of frames used in the adjustments and limiting the increase of $N/S$ at short wavelengths.

The RSCD step was not originally included in the pipeline. The first data reduction applied to the L168-9b data on 7 July 2022, corresponding to pipeline version 1.5.3, and CRDS context 0916 did not mention any RSCD step correction or frame-flagging steps related to the RSCD effect in the header. Prior to this work, reprocessed data with pipeline version 1.8.0 and CRDS context 1017 on 21 November 2022, did include a RSCD step in the header, but this step was not activated. This work is based on pipeline version 1.8.5 and CRDS context 1075, where a RSCD step does exist as well, but is still inactive. We activated it and demonstrated that the additional $N/S$ is now corrected when the related frames are flagged. However, activating this step does not consist in a proper correction of the additional $N/S$. To do this, a more robust algorithm for long-duration time-series data, especially for very bright targets with a small number of groups per integration, would use the temporal axis to detect outliers in the integration ramp for a given pixel. A method like this is in fact implemented in the JWST calibration pipeline as of version 1.11, and reprocessing the data with this updated jump detection algorithm has confirmed our hypothesis for the root cause of the excess $N/S$. The update now allows us to achieve the excellent noise properties predicted by the simulations, without having to discard any groups from the dataset. In comparison to the simulations, the RSCD effect was not applied to the ramp, and therefore, no excess of $N/S$ was witnessed at short wavelengths.

This additional $N/S$ has not been seen in any other LRS dataset to date, including targets dominated by photon noise, such as WASP-43, GJ1214, and WASP-107 observations from program IDs 1366, 1803, and 1280 \citep{bell_first_2023, kempton_reflective_2023}, and calibration target BD+60-1753, program ID 1053. The main difference between these targets and L168-9 is the apparent magnitude. L168-9 is much brighter, and it takes only nine frames before the detector is saturated. In comparison, the other targets display a large number of frames, from 40 to more than 100. Our work raised the question whether bright targets are associated with a low number of frames. This requires particular attention in terms of data reduction.

\section{Investigating persistence effects}
\label{section:persistence-effect}

In this section, we present a comprehensive investigation of the persistence effects found in the L168-9b data based on a quantitative analysis of their amplitude, time constant, and flux dependence. As presented in Sect.~\ref{section:MIRISIM_TSO}, persistence effects are a deficit or a surplus of signal at the beginning of a time-series observation, and they are due to previous operations of the detector. In order to quantify the impact of these effects on exoplanet time series observed with the MIRI LRS slitless mode, we computed the light curve of each pixel of the slitless subarray. Depending on the location of the pixel in the subarray and therefore on the amount of flux it receives, we witness different types of persistence effects in the data. 

\begin{figure*}[ht!]
    \centering
    \includegraphics[width=0.6\paperwidth]{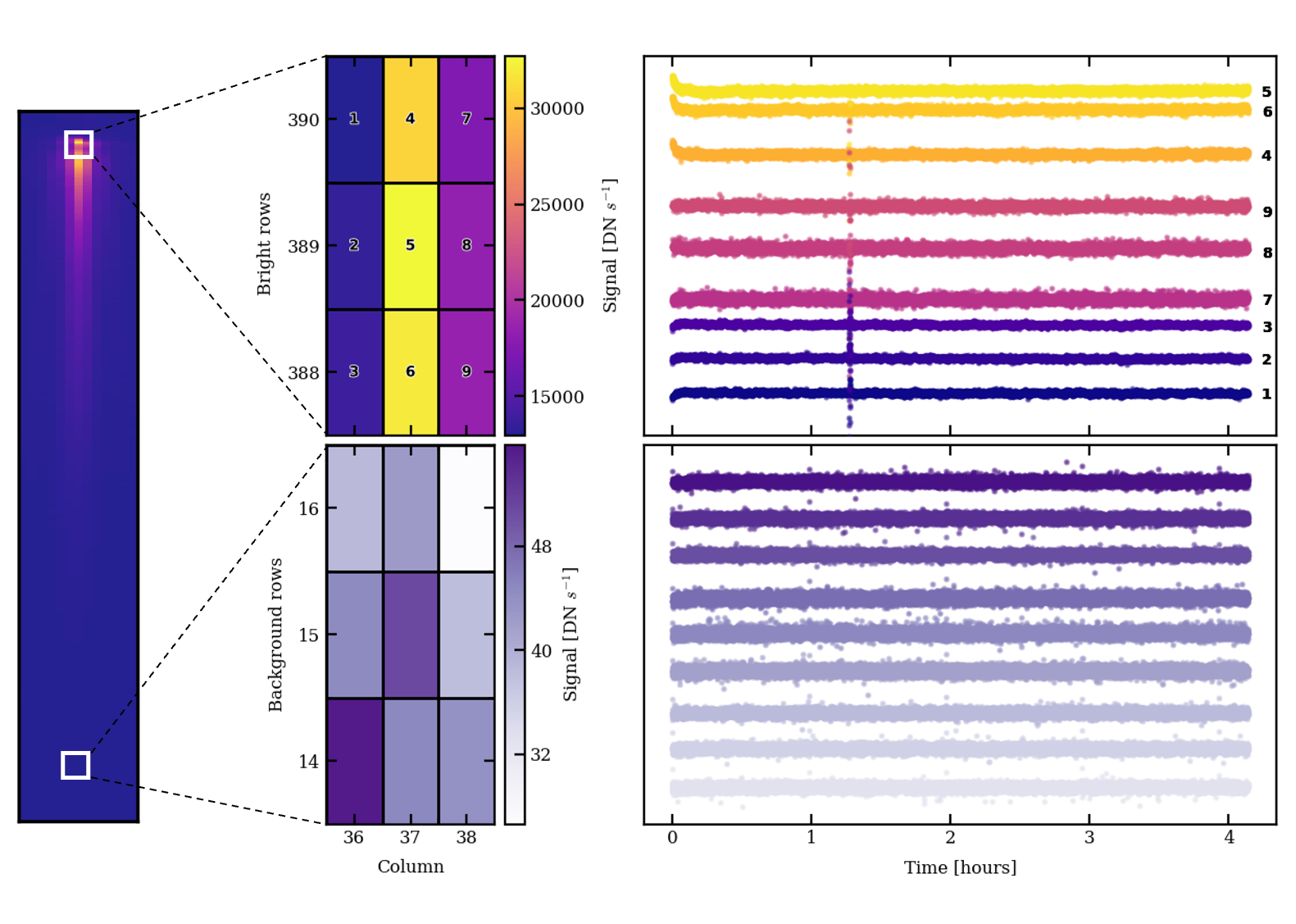}
    \includegraphics[width=0.4\paperwidth]{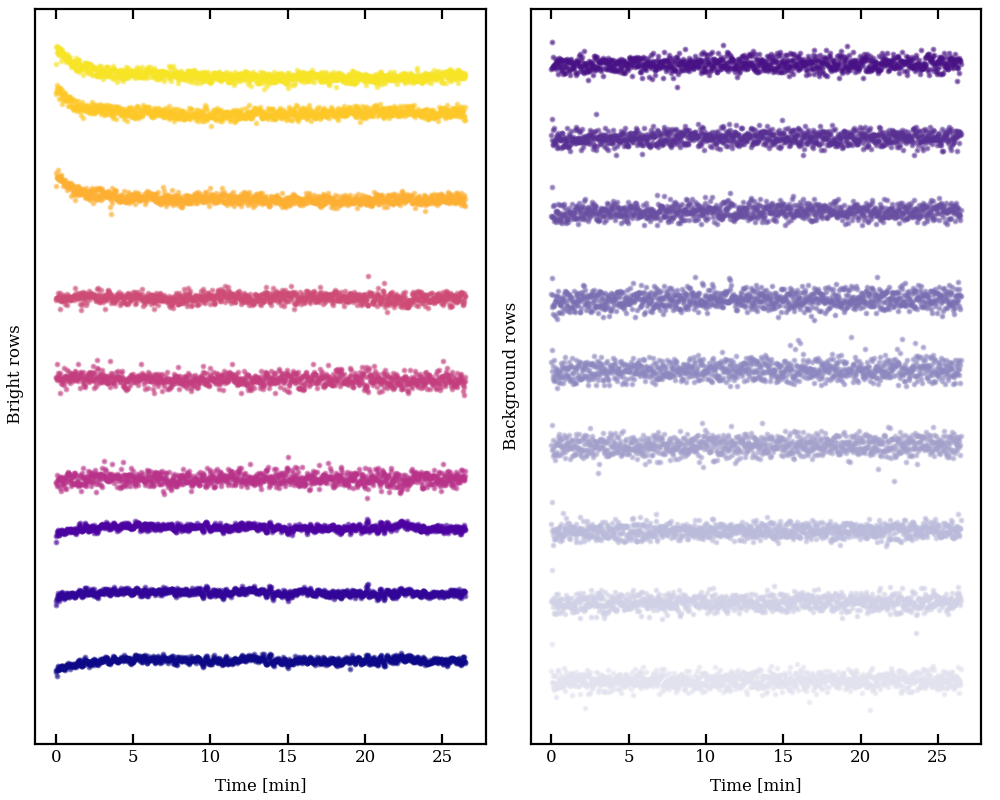}
    \caption{Depiction of persistence effects in the L168-9b data. Upper part: Full light curves. Bottom part: Zoom into the beginning of the light curves up to half an hour. \textit{Top left panel:} Rectangular selection of nine pixels located in the spectral trace, at the highest flux levels. \textit{Top right panel:} Corresponding time series. Persistence effects are visible at the beginning of each time series for less than approximately 30 minutes. \textit{Bottom left panel:} Selection of pixels located in the background. All pixels display a really low but yet similar amount of flux between 30 and 50 $\mathrm{DN} \: \mathrm{s}^{-1}$. \textit{Bottom right panel:} Corresponding time series in which no persistence effect is visible for these pixels.}
    \label{fig:persistence_exemples}
\end{figure*}

Fig.~\ref{fig:persistence_exemples} shows the light curves of a set of pixels taken from the slitless subarray. The top left panel shows a rectangular selection of nine pixels located in the spectral trace, at the highest flux levels. The top right panel shows the corresponding time series. Pixel 5, and therefore, the column amid the selection (depicted in yellow), is the brightest pixel in the subarray, with a flux level up to 32 000 $\mathrm{DN} \: \mathrm{s}^{-1}$. The columns on the right (in light purple) and on the left (in dark purple) receive 38 \% and 57 \% less flux on average, respectively, than the columns in the middle. The difference in flux between the column on the left and on the right points out an asymmetry of the PSF, which is centred on columns 37 and 38. The loss of signal at 1.2 hours of observation, which is visible in the light curves, is due to the HGA move during the exposure. The bottom left panel shows a selection of pixels located in the background. All pixels display a very low, but still similar amount of flux between 30 and 50 $\mathrm{DN} \: \mathrm{s}^{-1}$. The corresponding time series are displayed in the bottom right panel. Remarkably, no significant persistence effect is visible for these pixels.

Persistence effects are only visible in the top right panel at the beginning of each time series, for less than approximately 30 minutes. Each column shows a different behaviour of these persistence effects. The light curves of pixels located in the brightest column display a decay of flux, whereas those from the left column display an increase in flux over time. Those from the right column display no change in flux at the beginning of the observation. Those from the background do not seem to show any persistence effect at all. 
The discrepancies between the persistence effects show that they are highly flux dependent. We therefore point out several flux regimes. First, high fluxes above 30 000 $\mathrm{DN} \: \mathrm{s}^{-1}$ are impacted by the idle effect, introduced in Sect.~\ref{section:MIRISIM_TSO}. Then, as the LRS slitless subarray receives background flux even when no target is observed, the response drift has less impact than during a first complete illumination. Therefore, fluxes between 15 000 and 30 000 $\mathrm{DN} \: \mathrm{s}^{-1}$ must be in an equilibrium state between idle and response drift. Fluxes lower than 15 000 $\mathrm{DN} \: \mathrm{s}^{-1}$ seem to be affected by the response drift effect, which occurs when the subarray is illuminated after a pause in the observation. 
Finally, fluxes with an order of magnitude of a few tens, generally coming from the observational background, are only affected by a very low amplitude response drift. 

The overall aspect of the light curves after spectral binning between 5 and 12 $\mu$m, displayed in Fig.~\ref{fig:eureka_compa}, mainly shows a decay rather than an increase in flux, except at the highest wavelengths. The idle effect therefore dominates the response drift in the trace. As discussed in Sect.~\ref{section:l168_simulations}, this justifies the use of the idle effect alone in our simulations.

\section{Perspectives}
\label{section:futur_dev}

\subsection{Test of several atmospheric models}
\label{section:test_atmosphere}
 
In Sect.~\ref{data_red_analysis} we have demonstrated that we are able to produce realistic MIRI LRS slitless simulations. 
In this section, we simulate MIRI LRS data for two atmospheric scenarios for L168-9b:
\begin{itemize}
    \item Scenario 1: a hydrogen- and helium-rich atmosphere.
    \item Scenario 2: a thick Venus-like atmosphere with a metallicity X 1000, 50\% $\rm CO_2$, and 15\% $\rm CO$
\end{itemize}

The atmospheric absorption model was generated using \texttt{ATMO}, a 1D-2D radiative-convective equilibrium model for planetary atmospheres \citep{tremblin_fingering_2015,tremblin_cloudless_2017}. It solves the radiative transfer equation for a given set of opacities and computes a P-T profile that satisfies hydrostatic equilibrium and conservation of energy. It can compute equilibrium and non-equilibrium chemical abundances with the kinetic network of \cite{venot_chemical_2012}. Here, for the two atmospheric scenarios, we assumed an equilibrium chemistry. 

\begin{figure}[ht!]
    \centering
    \includegraphics[width=0.8\columnwidth]{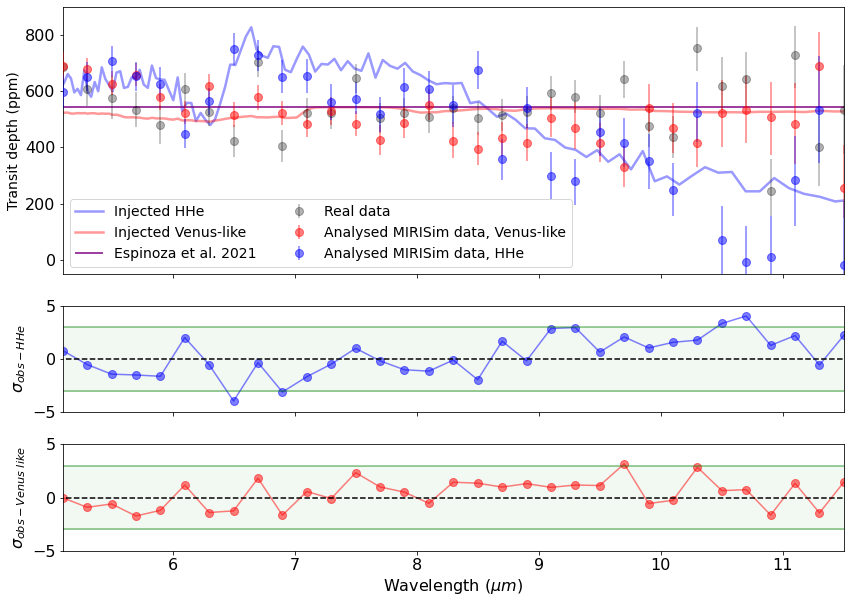}
    
    \caption{Transmission spectrum of L168-9b from the analysis of simulated data with \texttt{MIRISim-TSO}. {\it Top:} Transmission spectra for two atmospheric scenarios, Venus-like (red dots) and H/He-dominated (blue dots), as well as from real commissioning data (grey dots). {\it Middle:} Differences between the observed transit depth and the simulated transit depth for a hydrogen-helium-dominated atmosphere, where $\sigma = \frac{(depth_{obs}[\lambda]-depth_{HHe}[\lambda])}{\sqrt{err_{depth,obs}^{2}+err_{depth,HHe}^{2}}}$, 3$\sigma$ range indicated in green. {\it Bottom:} Same as middle panel, but for the observed transit depth and the simulated transit depth for a Venus-like atmosphere. }
    \label{compa_atmos_models}
\end{figure}

Similarly to Sec. \ref{section:l168_simulations}, we used \texttt{exoNoodle} and \texttt{MIRISim-TSO} to create two sets of MIRI LRS slitless simulations for each atmospheric scenarios, which we then analysed using \texttt{Eureka!}, as described in section \ref{section:reduction_3_5}. Fig.~\ref{compa_atmos_models} shows the injected and retrieved transmission spectra for the two scenarios. We confirm that we retrieved the expected molecular features in the case of an H/He-rich atmosphere. To assess which simulated data set fits the real observations best, we computed the $\chi^{2}$ statistic for the two simulated scenarios and for an airless planet with radius $R_p = 1.39 R_\odot$ \citep{astudillo-defru_hot_2020}. We obtained a $\chi^2$ of 328, 126, and 61 for a H/He-rich atmosphere, a Venus-like atmosphere, and an airless planet, respectively. The data are therefore best fit with an airless model, and a hydrogen-rich atmosphere is the most unlikely scenario. This result is consistent with the result obtained from previous studies \citep{astudillo-defru_hot_2020,bouwman_spectroscopic_2022}. The current spectrophotometric precision is insufficient to distinguish between an airless planet and a high mean-molecular weight thick atmosphere scenario for L168-9b, however.


\subsection{Future work on the data reduction of bright targets}

The impact of the RSCD effect, depicted in Sect.~\ref{section:nsr}, on the additional $N/S$ of L168-9b data is still under investigation, and it is a prominent question whether this effect could impact future observations. An additional $N/S$ has a direct consequence on the transmission feature error bars and on the baseline that we obtain after data reduction. One possible way of investigating this is to work on up-the-ramp non-linearities that could impact the ramp-fitting step. The ramp is fitted by a least-squares minimisation algorithm \citep{robberto_generalized_2013} that provides the slope value and its variance estimate. A high variance could lead to higher error bars, and therefore, to an additional $N/S$. Before fitting the ramps, a non-linearity correction step is applied. This step includes cosmic-ray correction and an up-the-ramp non-linearities correction within a 5\% error margin specification \citep{rieke_infrared_2007, rieke_mid-infrared_2015}. 
The ramp-fitting of bright targets may become more critical as the fit is performed on only a few frames. Moreover, if the signal level in a pixel exceeds 10 000 DN, the last frames are likely to be impacted by the brighter-fatter effect that introduces non-linearities as well, which alters not only the signal level, but also the size of the PSF. The brighter-fatter effect \citep{argyriou_brighter-fatter_2023} is caused by charge migration from the central brightest pixel of the PSF into the surrounding pixels as charge accumulates towards saturation. The effect causes a broadening of the PSF, and it is particularly marked at short wavelengths, where the contrast between the central and nearby pixels is stronger; a secondary manifestation is a steepening of the ramp in the pixels into which additional charge is migrating, which adds to the residual non-linearities in their ramps. As presented by \cite{morrison_jwst_2023}, the RSCD effect also includes non-linearities at the beginning of the ramp. A correction for the RSCD effect may therefore be needed to remove the resulting non-linearity behaviour and to keep the impacted frames in the dataset to provide more frames for the fit. Thus, correcting for all non-linearities in general may become crucial to provide a linear ramp to fit and to keep all the frames within the dataset. Finally, the next step is to examine all the LRS slitless observations that are available to date, including calibration data acquired during commissioning on HD167060, HD180609, and HD37962, to derive correlations between the non-linearity behaviour and the number of frames, as well as correlations between non-linearities and the signal and flux levels in a pixel. We expect the outcome of this work to provide key insights into the extent of non-linearities and correction methods that are currently at stake.

\subsection{Future work on persistence effects}

As presented in Sec.~\ref{section:persistence-effect}, the persistence effects witnessed in the L168-9b data are flux-dependent and affect the spectral time series. The only way to remove them from the dataset is to fit them with a model made of one or two exponentials or even with a polynomial function. As persistence effects occur at the beginning of an exposure, they can be easily fitted or removed from transit or eclipse observations. As shown in \cite{bouwman_spectroscopic_2022}, the residuals of the L168-9b transit fitting do not show any excess of red noise, as shown in Fig.~\ref{fig:Allan_5_7}. However, phase-curve modulations are strongly correlated to persistence effects, and fitting them is therefore highly degenerate. \cite{bell_first_2023} showed the strong impact of persistence effects in the MIRI LRS slitless phase-curve observation of the ERS target WASP-43b. Not only are they strongly correlated to the phase-curve parameters, they also follow the same regimes of the idle effect and response drift as found in the L168-9b data in Sect.~\ref{section:persistence-effect}. Furthermore, \cite{bell_first_2023} showed that persistence effects are not only flux dependent, but also result from the pixel location in the subarray. As the LRS slitless subarray overlaps the MIRI coronographs \citep{boccaletti_mid-infrared_2015} and constantly receives observational background, the persistence effect may depend on the filter position prior to the observation. Long-wavelength filters are likely to increase the observational background. To characterise the persistence effects, one way is to pull down the telemetry of the MIRI LRS slitless subarray to find correlations between these effects and the filter wheel position and the time spent idling prior to the observation.


\section{Conclusions}

We introduced realistic simulations of transiting exoplanets with the MIRI LRS slitless mode. In particular, we simulated the observation of L168-9b, a super-Earth-sized exoplanet chosen to meet the requirements of the instrumental stability calibration program conducted during commissioning in the LRS mode of the MIRI instrument. To ensure that our simulations complied with real data, we refined and adapted the detector set-up, the systematics, and the persistence effects, thus conforming to the in-flight calibration. Finally, we demonstrated that simulations replicate the data with similar scatter and error bars.

Our simulations provide key insights for the understanding of instrumental systematics of MIRI LRS. First, we established that activating the RSCD step that flags the first four frames in the \texttt{jwst} Stage 1 is the way to eliminate the additional noise-to-signal ratio at short wavelengths, which was originally witnessed by \cite{bouwman_spectroscopic_2022}. Then, our work demonstrated that residual non-linearities remaining at the ramp level lead to an incorrect cosmic-ray rejection, mostly for pixels located at short wavelengths.
A more robust algorithm for the cosmic-ray rejection step, especially for very bright targets with a small number of groups per integration, has very recently been implemented in the pipeline (version 1.11). The update now allows us to achieve the excellent noise properties predicted by the simulations, without having to discard any frames from the dataset.

This paper also agrees in terms of the persistence effects with the first results by \cite{bell_first_2023} on the phase-curve observation of WASP-43b. Our work shows two regimes of persistence effects in the L168-9b data, the idle effect and the response drift. Although persistence effects in the WASP-43b data show a higher amplitude and take more time to vanish, their structure is the same as that of those witnessed in the L168-9b data. A pioneering work to understand these effects and correct them has to be conducted to ensure reliable spectral retrievals without any degeneracy with the phase-curve models.

Most of all, we validated the data reduction and data analysis methods as we are able to retrieve a posterior distribution of parameters centred on the same values as were injected in our simulations. We notably observed that an H-He dominated atmosphere is the least favoured scenario ($\chi^2$ = 326) compared to a Venus-like atmosphere scenario ($\chi^2$ = 133) and an airless scenario ($\chi^2$ = 61).

\begin{acknowledgements}
    E. D acknowledges support from the innovation and research Horizon 2020 program in the context of the  Marie Sklodowska-Curie subvention 945298. 
    PT would like to acknowledge and thank the ERC for funding this work under the Horizon 2020 program project ATMO (ID: 757858).
    P.O.L. acknowledges funding support from CNES. This work uses observations made with the NASA/ESA/CSA James Webb Space Telescope. The data were obtained from the Mikulski Archive for Space Telescopes at the Space Telescope Science Institute, which is operated by the Association of Universities for Research in Astronomy, Inc., under NASA contract NAS 5-03127 for JWST. These observations are associated with calibration program 1033. This work is based [in part] on observations made with the NASA/ESA/CSA James Webb Space Telescope. The data were obtained from the Mikulski Archive for Space Telescopes at the Space Telescope Science Institute, which is operated by the Association of Universities for Research in Astronomy, Inc., under NASA contract NAS 5-03127 for JWST. These observations are associated with program 1280. MIRI draws on the scientific and technical expertise of the following organisations: Ames Research Center, USA; Airbus Defence and Space, UK; CEA-Irfu, Saclay, France; Centre Spatial de Liège, Belgium; Consejo Superior de Investigaciones Científicas, Spain; Carl Zeiss Optronics, Germany; Chalmers University of Technology, Sweden; Danish Space Research Institute, Denmark; Dublin Institute for Advanced Studies, Ireland; European Space Agency, Netherlands; ETCA, Belgium; ETH Zurich, Switzerland; Goddard Space Flight Center, USA; Institut d’Astrophysique Spatiale, France; Instituto Nacional de Técnica Aeroespacial, Spain; Institute for Astronomy, Edinburgh, UK; Jet Propulsion Laboratory, USA; Laboratoire d’Astrophysique de Marseille (LAM), France; Leiden University, Netherlands; Lockheed Advanced Technology Center (USA); NOVA Opt-IR group at Dwingeloo, Netherlands; Northrop Grumman, USA; Max-Planck Institut für Astronomie (MPIA), Heidelberg, Germany; Laboratoire d’Études Spatiales et d’Instrumentation en Astrophysique (LESIA), France; Paul Scherrer Institut, Switzerland; Raytheon Vision Systems, USA; RUAG Aerospace, Switzerland; Rutherford Appleton Laboratory (RAL Space), UK; Space Telescope Science Institute, USA; ToegepastNatuurwetenschappelijk Onderzoek (TNO-TPD), Netherlands; UK Astronomy Technology Centre, UK; University College London, UK; University of Amsterdam, Netherlands; University of Arizona, USA; University of Bern, Switzerland; University of Cardiff, UK; University of Cologne, Germany; University of Ghent; University of Groningen, Netherlands; University of Leicester, UK; University of Leuven, Belgium; University of Stockholm, Sweden; Utah. We would like to thank the following National and International Funding Agencies for their support of the MIRI development: NASA; ESA; Belgian Science Policy Office; Centre Nationale d’Études Spatiales (CNES); Danish National Space Centre; Deutsches Zentrum fur Luft-und Raumfahrt (DLR); Enterprise Ireland; Ministerio De Economiá y Competividad; Netherlands Research School for Astronomy (NOVA); Netherlands Organisation for Scientific Research (NWO); Science and Technology Facilities Council; Swiss Space Office; Swedish National Space Board; and UK Space Agency.
    \textit{Software:} \texttt{numpy} \citep{harris_array_2020},  \texttt{matplotlib} \citep{hunter_matplotlib_2007}, \texttt{scipy} \citep{virtanen_scipy_2020}, \texttt{astropy} \citep{the_astropy_collaboration_astropy_2022}, \texttt{nuts} \citep{hoffman_no-u-turn_2011}, \texttt{emcee} \citep{foreman-mackey_emcee_2013}, \texttt{dynesty} \citep{speagle_dynesty_2020}, \texttt{starry} \citep{luger_starry_2019}, \texttt{exoNoodle} \citep{martin-lagarde_phase-curve_2020}, \texttt{MIRISim} \citep{klaassen_mirisim_2020}, \texttt{eureka!} \citep{bell_eureka_2022}, \texttt{pipeline\_parallel} (\url{https://gitlab.com/jwst_fr/pipeline_parallel/-/tree/master/}) and \texttt{jwst} (\url{https://jwst-pipeline.readthedocs.io/en/latest/}).
\end{acknowledgements}

\bibliographystyle{aa} 
\bibliography{references}
%
%


\appendix

\section{Description of the persistence effects}
\label{section:MIRISIM_TSO-description}

In the 2018 test data we obtained from JPL \citep{martin-lagarde_preparation_2020} of the MIRI detector, three persistence effects were identified: the response drift, the anneal recovery, and the idle recovery.

Response drift occurs when the detector is suddenly illuminated after a given time without receiving any flux. The output signal is not stabilised and slowly increases towards its expected value following a smooth slope \citep{rieke_mid-infrared_2015}. The stronger the flux in a pixel, the quicker the effect vanishes over time. 

Anneal recovery is induced by a previous heating of the detector. In an arsenic-doped silicon (Si:As) detector, such as the MIRI detector \citep{rieke_mid-infrared_2015}, collisions between charged particles such as cosmic rays and silicon atoms may damage the crystal structure. This is called radiation damage \citep{oblakowska-mucha_radiation_2017}. Detector anneals are therefore performed to recover from this damage. The detector is heated up ~15 to 20 K higher than the nominal temperature and then cooled back down to 7 K. As a consequence, the output signal is higher at the beginning of an observation than the nominal value, and it smoothly decreases towards its expected value. This effect is called anneal recovery and is completely independent of the flux value in a pixel.

Idle recovery is the last persistence effect that is identified. Between two observing phases, the detector is still illuminated with the observational background. The longer the detector waits, the more background it acquires. To prevent the detector from acquiring any signal, consecutive resets are performed. This is called an idle \citep{argyriou_calibration_2021}. As soon as the observation starts, the output signal level is higher than expected because of these resets, and it slowly decreases towards its nominal value. This effect depends on the flux in a pixel and on the time during which idle has been performed. 

\section{L168-9b observation telemetry}
\label{section:telemetry}

\begin{figure*}[ht!]
    \centering
    \includegraphics[width=88mm]{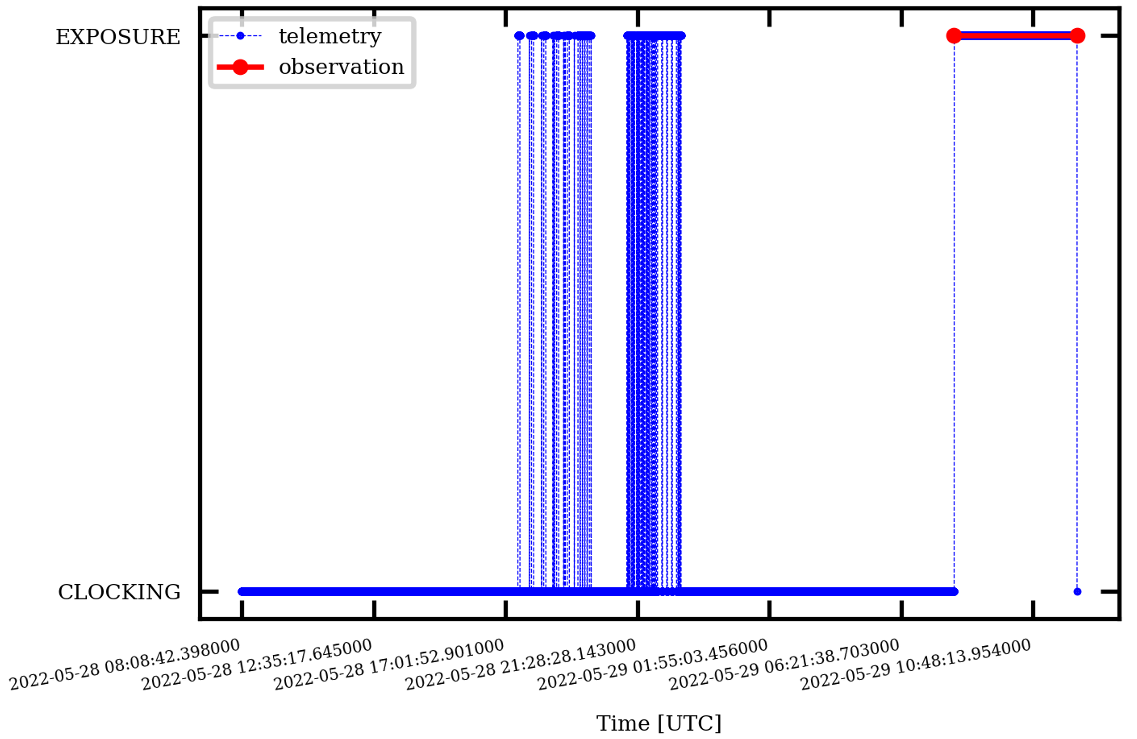}
    \caption{MIRI telemetry during the 48 hours preceding the observation of L168-9b that provides the detector status, either in exposure (observing) mode, or in clocking mode (in repeated idle procedure). The observation is preceded by a series of idles lasting 9h13.}
    \label{fig:telemetry}
\end{figure*}

The JWST telemetry of the MIRI detector status was pulled to obtain the detector status before the observation and thus characterise the idle effect witnessed at the beginning of the observation (see Fig.~\ref{fig:white_data}). The status is either EXPOSURE, which means that the detector is observing (receiving light from a source or the background), or CLOCKING, which means that the detector is resetting continuously (i.e. idling). This curve was obtained by pulling the telemetry under mnemonic IMIR\_IC\_ACTIVE\_DET\_MODE. Fig.~\ref{fig:telemetry} shows the time spent idling (in blue) before the observation (depicted in red in the figure), which is about 9 hours. 

\section{Correlated noise computation}
\label{section:correlated_noise}

\begin{figure*}[ht!]
    \sidecaption
    \includegraphics[width=12cm]{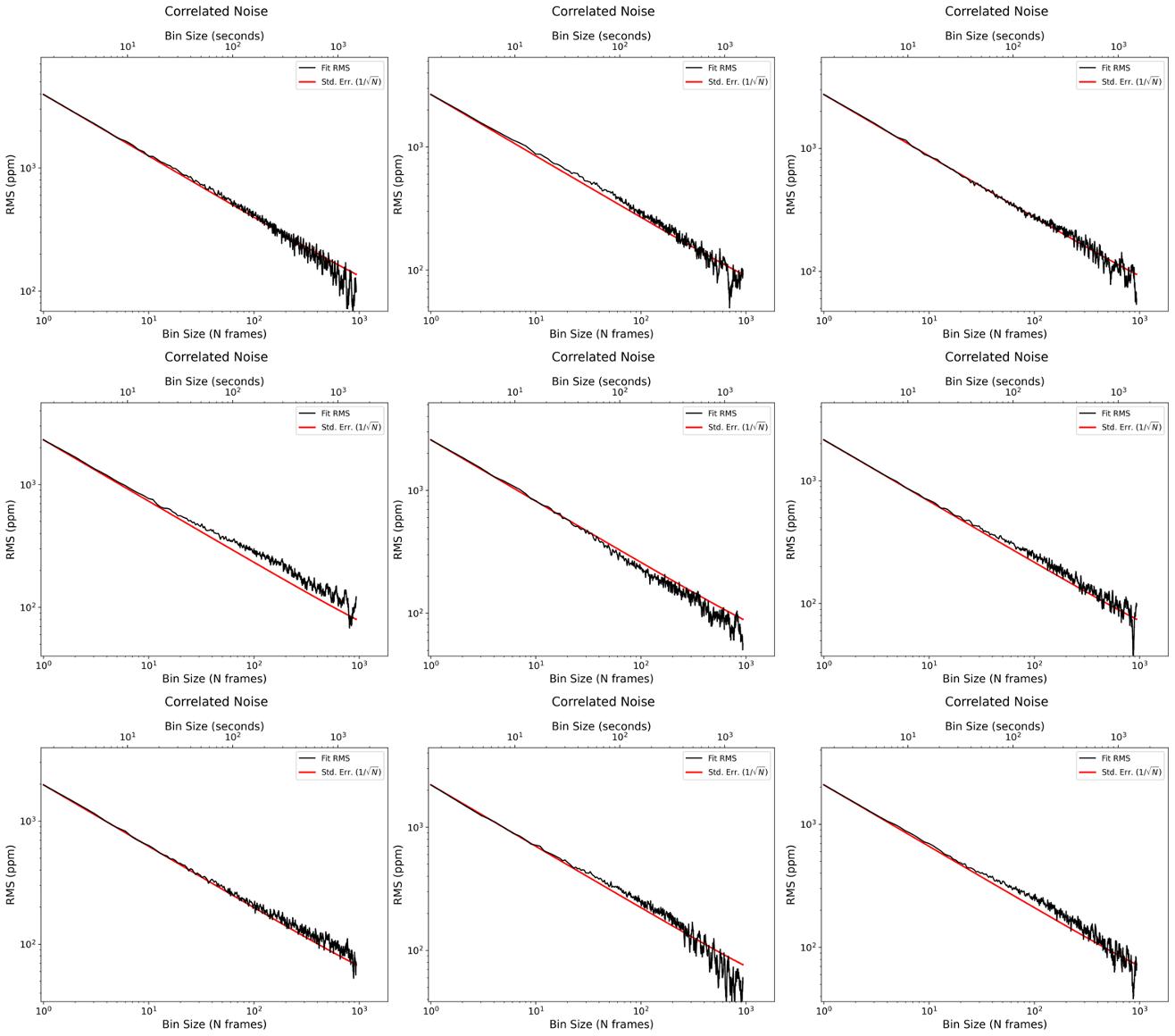}
    \caption{Allan plots of the first nine spectral bins of the L168-9b data between 5 and 7 $\mu$m. All plots show the minimum correlated noise in the residuals of the fit.}
    \label{fig:Allan_5_7}
\end{figure*}

The L168-9b data fit shows the minimum correlated noise in the residuals of the white light-curve as well as in the residuals of all spectroscopic light curves. A selection of the first nine plots of the Allan variance computation is displayed in Fig.~\ref{fig:Allan_5_7}. The  first nine channels refer to the wavelength range between 5 and 7 $\mu$m, where the excess of noise-over-signal is presented in \cite{bouwman_spectroscopic_2022}.

\section{Alternative method for the ramp fitting}
\label{section:no_ramp_fit_method}

Starting from the L168-9b uncalibrated spectral images, we applied an alternative method to the data reduction:
\begin{enumerate}
\item First, each ramp starts with an offset value that has to be removed from all ramps of the spectrum. To do this, we chose a region in the lower part of the trace that corresponds to the background zone. We chose a rectangular selection from row 0 to row 50 and from columns 35 to 38. For each ramp, we selected the penultimate frame, and we computed the mean over the whole region of this signal level. We did not use the last frame because it is impacted by the last-frame effect, which is strongly non-linear. 
\item The second step is to obtain the signal over the spectrum. We selected a region from row 155 to row 391 and from column 35 to column 38, which corresponds to the trace. For each row, we summed the signal over the three columns to obtain a unique value of the signal per row. Then, we chose the penultimate frame of each ramp and subtracted the offset value. In this way, each row was assigned a unique signal value that evolved over time.
\item Then, each light curve was cleaned by applying a temporal sigma clipping with a running mean of 50 values and a rejection threshold of 5$\sigma$.
\item The fourth step is to derive the mean value of the signal over time, only taking into account the stable part of the observation, after integration 8900. This step provides a spectrum as a function of the detector row presented in Fig~\ref{fig:signal_new}. 
\begin{figure}[ht!]
    \centering
    \includegraphics[width=1\columnwidth]{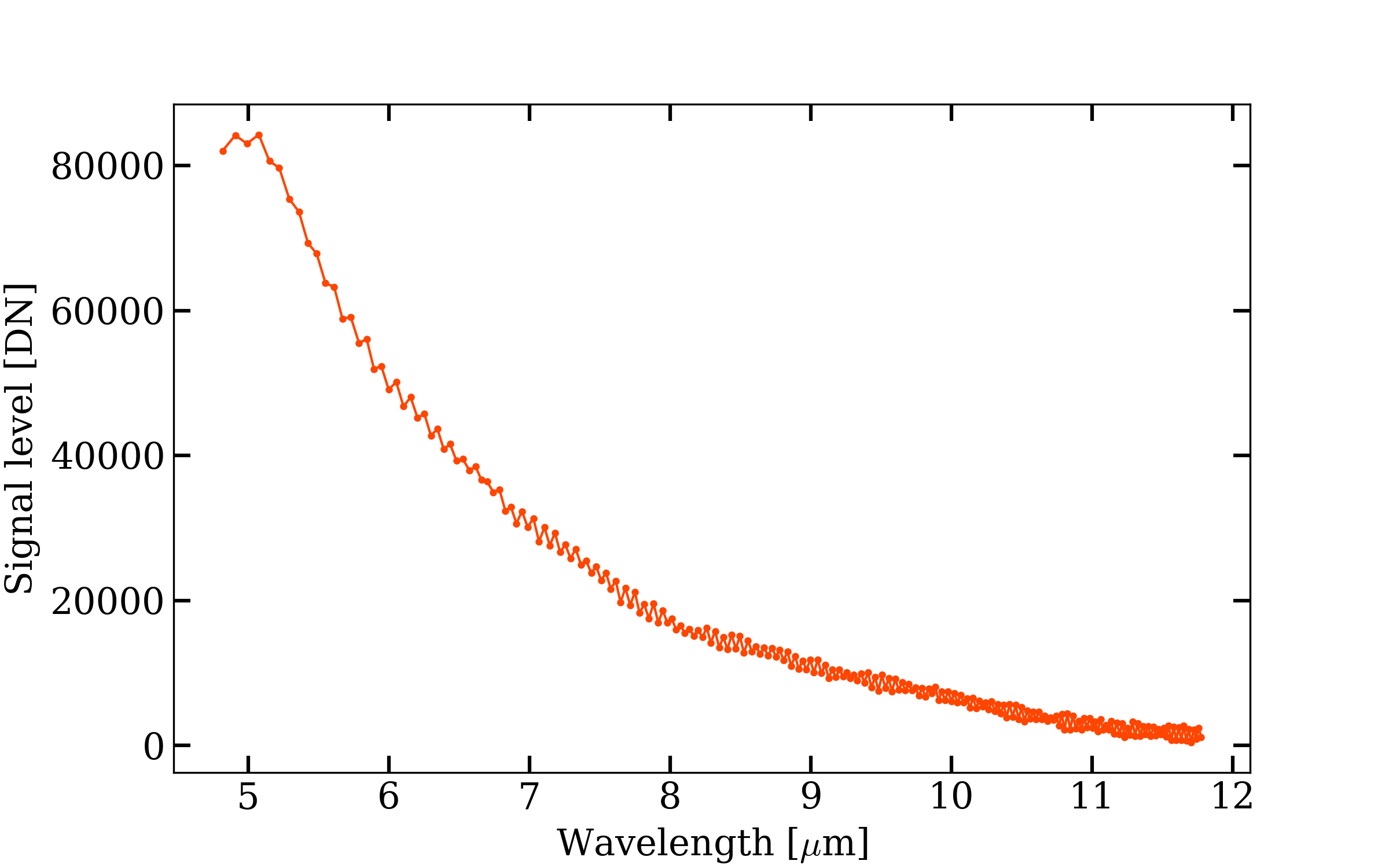}
    \caption{L168-9b spectrum in DN extracted using only the penultimate frame of the ramp. No ramp-fitting is applied.}
    \label{fig:signal_new}
\end{figure}
\item The next step is to evaluate the noise. This was done by estimating the standard deviation of each light curve. To remove outliers, a spectral sigma clipping was applied with a running mean of ten rows and a rejection threshold of 5$\sigma$. The outcome of this step is presented in Fig~\ref{fig:noise_new}.
\begin{figure}[ht!]
    \centering
    \includegraphics[width=1\columnwidth]{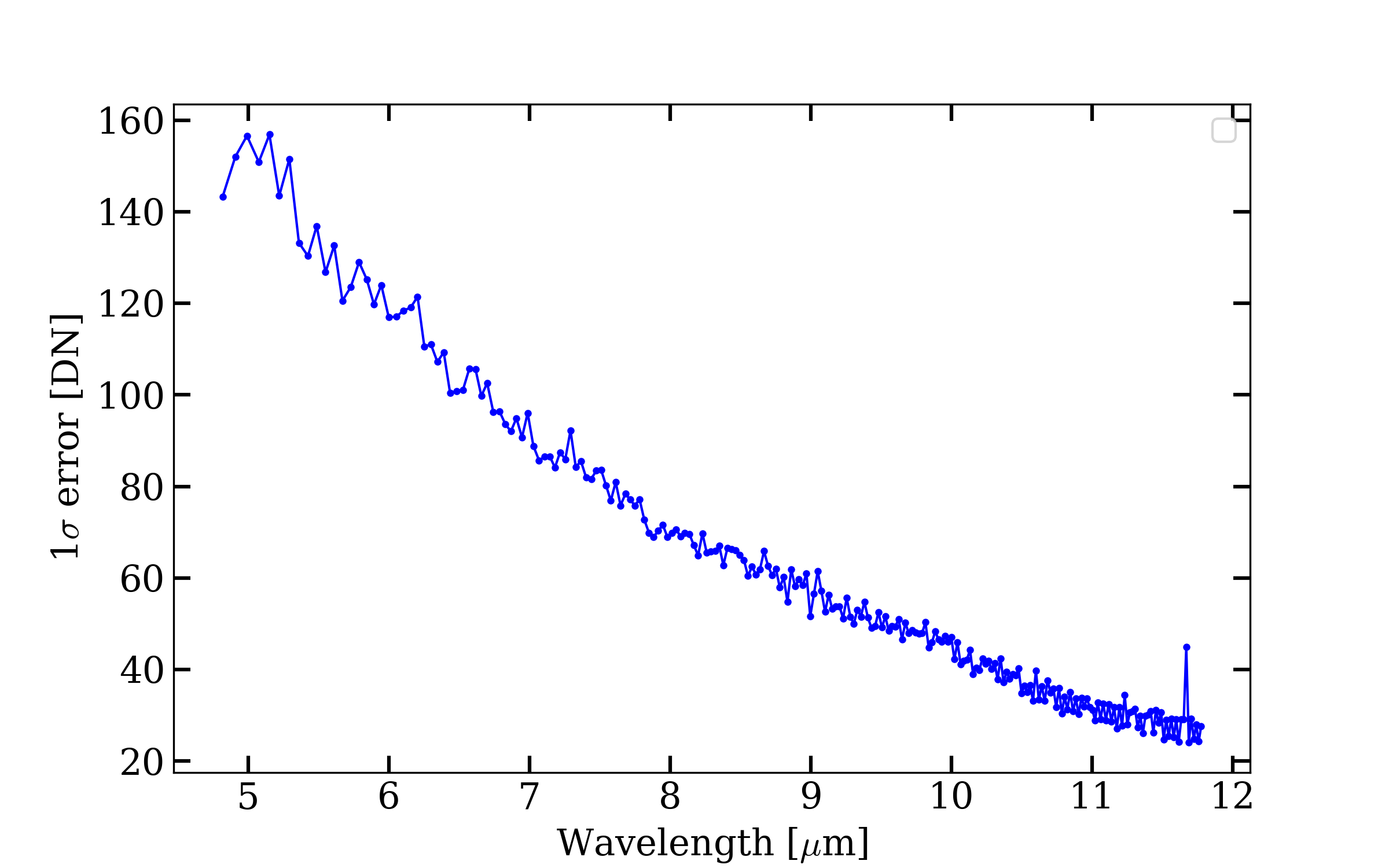}
    \caption{Noise estimate in DN of the L168-9b observation extracted using only the penultimate frame of the ramp.}
    \label{fig:noise_new}
\end{figure}
\item The final step is to divide the noise by the signal to obtain the noise-over-signal estimate as a function of the row. To present a proper result, we used the pixel-to-wavelength dispersion file made during commissioning. We obtained the noise-over-signal estimate as a function of wavelength in this way. 
\end{enumerate}

\begin{figure*}[ht!]
    \centering
    \includegraphics[width=88mm]{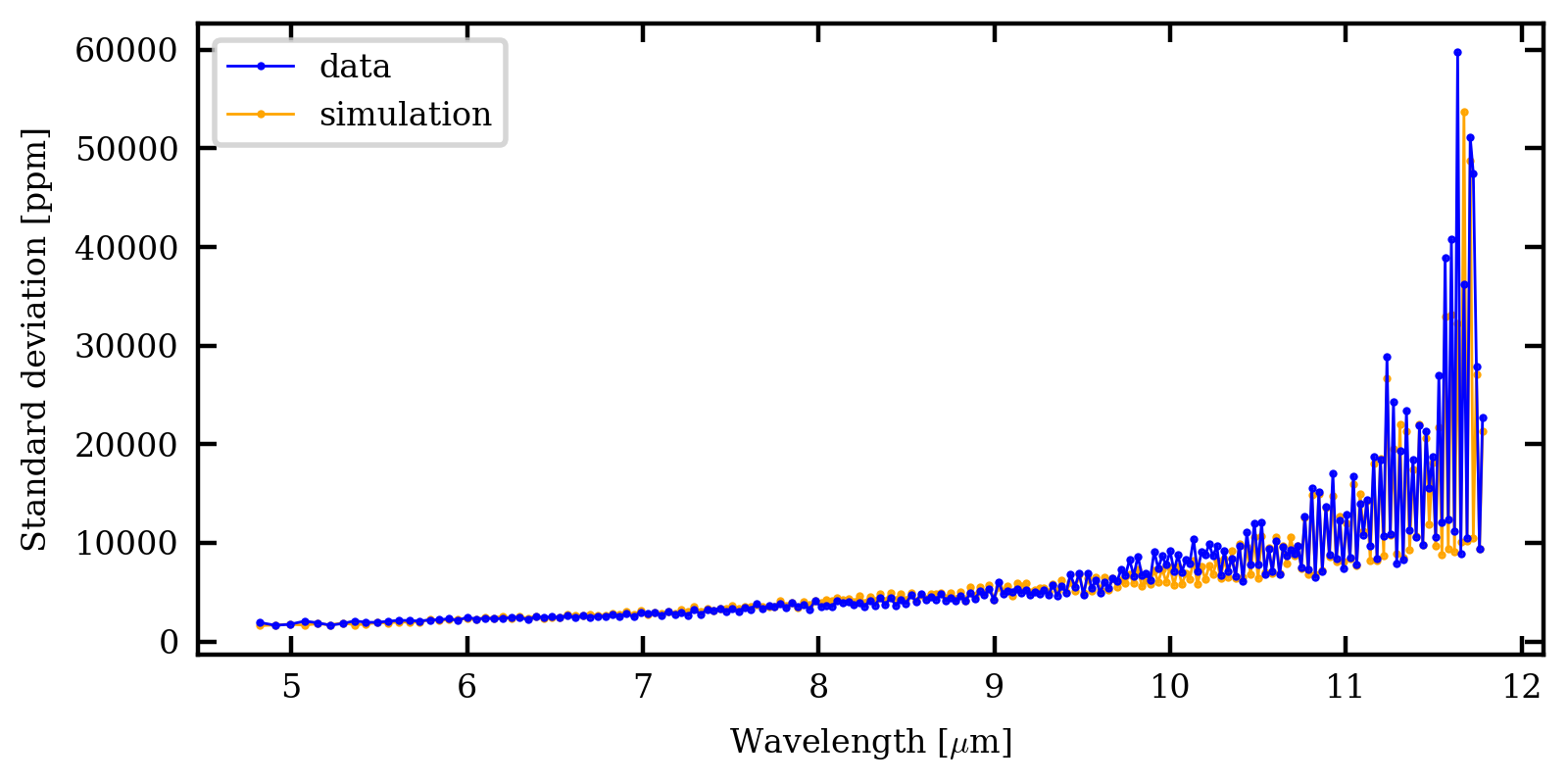}
    \includegraphics[width=88mm]{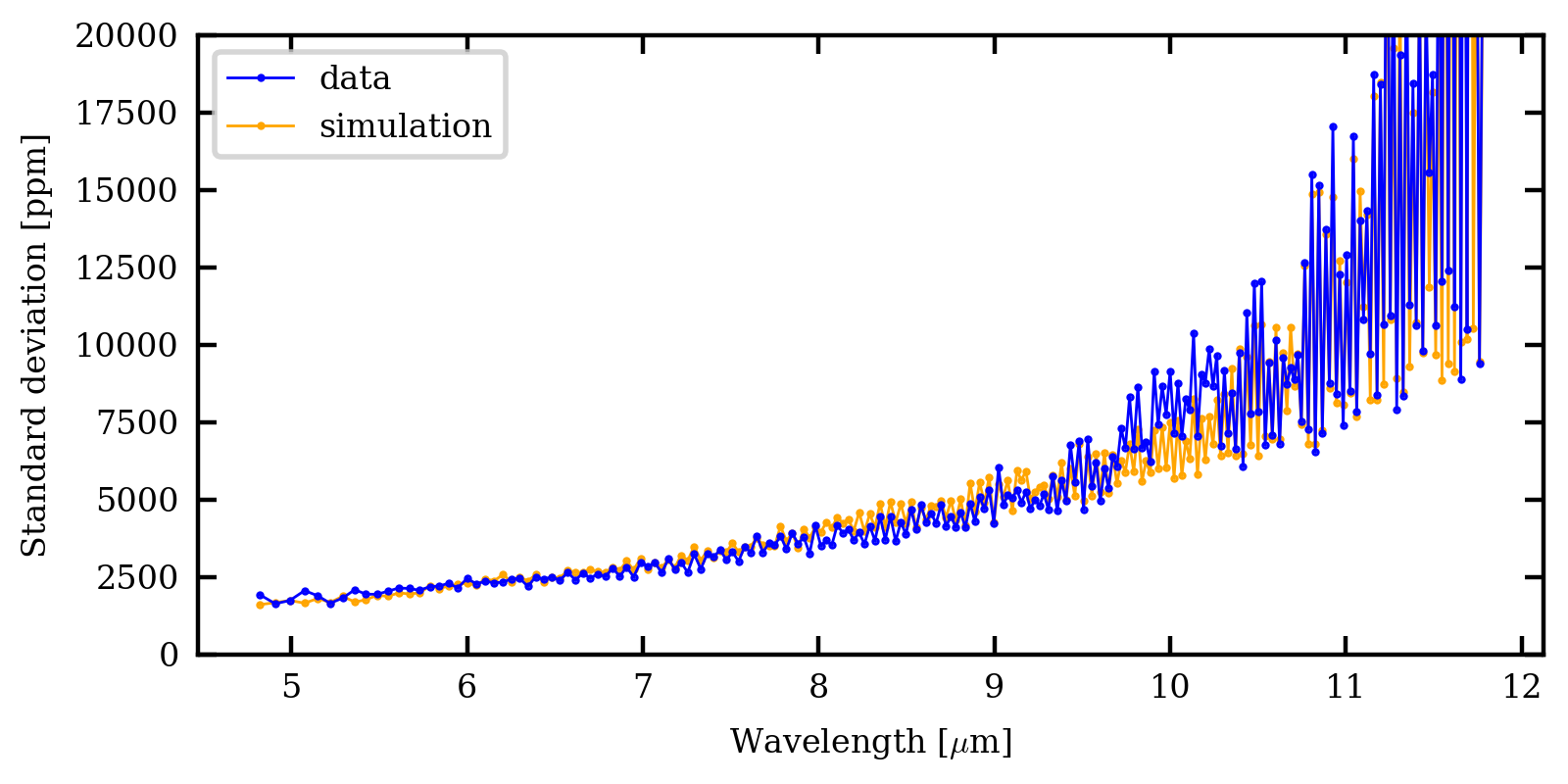}
    \caption{\textit{Top:} Noise-over-signal ($N/S$) estimate for the L168-9b data between 5 and 12 \si{\micro\meter}, using only the penultimate frame of the ramp to compute the signal, instead of fitting the ramp. The blue curve shows the data $N/S$ estimate, and the orange curve shows the simulation $N/S$ estimate. \textit{Bottom:} Zoom-in between 2500 and 20000 ppm to focus on the short wavelength $N/S$ values.}
    \label{fig:NSR_new_method}
\end{figure*}

The results are displayed in Fig~\ref{fig:NSR_new_method} for the real data and simulations. 

\end{document}